# Cosmological Formation of Low-Mass Objects

Zoltán Haiman[1,3], Anne A. Thoul[2], and Abraham Loeb[1,2]

## ABSTRACT

We investigate the early formation of bound objects with masses comparable to the cosmological Jeans mass ($\sim 10^5 M_\odot$). We follow the growth of isolated spherically symmetric density peaks starting from the linear perturbative regime. The initial parameters correspond to density peaks of various widths and heights in a cold dark matter cosmology. We use a one-dimensional spherical Lagrangian hydrodynamics code to follow the dynamical, thermal, and non-equilibrium chemical evolution of the gas. The system includes a collisionless dark matter component and a baryonic component composed of the nine species $\mathrm{H}, \mathrm{H}^-, \mathrm{H}^+, \mathrm{He}, \mathrm{He}^+, \mathrm{He}^{++}, \mathrm{H}_2, \mathrm{H}_2^+$, and $\mathrm{e}^-$. All relevant chemical reactions between these species and their cooling mechanisms are included in the calculations.

We explore the dependence of the dynamical evolution of the gas on two parameters: the initial mass scale and the initial overdensity of the system. We follow the evolution of the density, temperature, and abundance profiles within the cloud, assuming two types of central boundary conditions for the collisionless component: in one the infalling dark matter virializes through a reflection from a hard sphere, while in the other it accretes onto a central sink. We find that in both cases, radiative cooling by $\mathrm{H}_2$ affects the collapse dynamics of the gas only after it has already virialized and become part of the bound object. Therefore, radiative cooling is unlikely to have triggered the initial collapse of perturbations at redshifts $z > 10$. Nevertheless, baryonic objects with masses well below the linear-theory Jeans mass ( $\lesssim 10^3 M_\odot$) form at high redshifts due to shell crossing by the dark matter. Such objects could be the progenitors of a primordial population of high-mass stars in the intergalactic medium.

---

[1]Astronomy Department, Harvard University, 60 Garden Street, Cambridge, MA 02138, USA

[2]Institute for Advanced Study, Olden Lane, Princeton, NJ 08540, USA

[3]Institute of Astronomy, Cambridge University, Madingley Road, Cambridge CB3 1RZ, England

## 1. INTRODUCTION

Observations indicate that by a redshift $\sim 5$ the universe contained nonlinear objects such as quasars and absorption systems (see, e.g. review by Woltjer 1990). As current quasar searches are limited from probing even higher redshifts by practical considerations, there is no reason to suspect that they have already revealed the first generation of baryonic objects in the universe. In fact, popular Cold Dark Matter (CDM) cosmologies predict the collapse of the first baryonic objects as early as $z \sim 30$ on a mass-scale just above the cosmological Jeans mass $\sim 10^5 M_\odot$ (Peebles 1993).

The earliest baryonic objects are interesting for a variety of reasons. First, in the bottom-up hierarchical picture of structure formation, they provide the elementary building blocks for larger mass objects that form later. For example, Peebles & Dicke (1968) have suggested that globular clusters may be the fossils of Jeans mass objects that collapsed and fragmented into stars at high redshifts, long before galactic halos formed. Second, direct detection of low-mass objects at $z \gtrsim 10$ can be used to place constraints on the power-spectrum of primordial density fluctuations on very small scales that cannot be probed by microwave background anisotropy experiments (e.g., White et al. 1994). Third, the first baryonic objects may have fragmented into stars and contaminated the intergalactic medium with metals through supernova-driven winds. Recent spectroscopic observations with the Keck telescope indicate the existence of nonzero metalicity even in Ly$\alpha$ clouds (Fan 1994; Tytler & Fan 1995).

The evolution of the first baryonic objects depends on their molecular composition and thermodynamic state. The chemistry and radiative cooling during the cosmological collapse of these objects define a well-posed problem for investigation, since the element abundances are fixed by primordial nucleosynthesis and the cooling rates are not complicated (at least initially) by energy input or metal enrichment from stars. Under primordial conditions, molecular hydrogen ($H_2$) could become the most efficient coolant of collapsing clouds. Indeed, it has been suggested by various authors (e.g. Saslaw & Zipoy 1967, Peebles & Dicke 1968, Hirasawa 1969, Matsuda et al. 1969, and more recently Palla et al. 1983, Lepp & Shull 1984, and Lahav 1986) that radiative $H_2$ cooling may, in fact, control the formation of the first generation of objects. However, these suggestions were based on highly simplified models involving homogeneous clouds that cool radiatively while undergoing a uniform free-fall collapse. In reality, a collapsing system becomes highly inhomogeneous as its central density increases. The resulting density profile tends to steepen until a shock develops (see, e.g. Bertschinger 1985 for self-similar solutions well above the Jeans mass). As the shock thermalizes the kinetic energy of the infalling gas, it alters considerably the thermal history of the gas relative to the predictions of the homogeneous collapse models.

In order to study the effect of molecular cooling on collapsing primordial clouds, it is necessary to follow both the dynamics of the gas as a whole and the chemical evolution of its various molecular components. Unfortunately, the full implementation of all relevant chemical reactions and cooling rates in three-dimensional hydrodynamic simulations (Cen & Ostriker 1994; Hernquist, Katz, & Weinberg 1995) requires an extensive computational effort and has not yet been accomplished. As the first step towards a future implementation of this type, we examine in this paper spherically symmetric systems. We use a one-dimensional Lagrangian code to follow the dynamical, thermal and chemical evolution of a spherical system made of collisionless dark matter and collisional gas.

The minimum baryonic mass of the first generation of objects is defined by the smallest scale over which the repulsive pressure force is unable to counteract the attractive gravitational force. Since molecular hydrogen formation becomes more effective as the gas density increases, $H_2$ cooling can provide a positive feedback that reduces the gas pressure as the density increases, thus allowing the density to increase even further. In this paper we address the question: *could molecular cooling lower the mass of the first baryonic objects down to low (possibly stellar) values?*

The outline of the paper is as follows. In § 2 we review the notion of a cosmological Jeans mass based on linear perturbation theory, and describe how the presence of dark matter, nonlinearities, or cooling, affect this notion. In § 3 we describe the numerical code and its ability to reproduce various analytic results. In § 4 we present numerical results for the evolution of clouds with different initial conditions. In particular, we identify the regime of mass-scales and density contrasts over which radiative cooling by $H_2$ affects the collapse dynamics. Finally, § 5 summarizes the implications of these results for the first generation of bound objects in the universe.

## 2. SPHERICAL COLLAPSE AND THE COSMOLOGICAL JEANS MASS

The Jeans length $\lambda_J$ was originally defined (Jeans 1928) in Newtonian gravity as the critical wavelength that separates oscillatory and exponentially growing sinusoidal density perturbations in an infinite, uniform, and stationary distribution of gas. On scales smaller than $\lambda_J$, the sound crossing time is shorter than the gravitational free-fall time, allowing the build-up of a pressure force to balance gravity. The Jeans mass is defined as the mass within a sphere of radius $\lambda_J/2$, $M_J = (4\pi/3)\rho(\lambda_J/2)^3$. In a perturbation with a mass greater than $M_J$, self-gravity cannot be supported by pressure, and so the gas is unstable to gravitational collapse. The Newtonian derivation of the Jeans instability suffers from a conceptual inconsistency, as the unperturbed gravitational force of the uniform background must induce bulk motions (cf. Binney & Tremaine 1987). However, this inconsistency is remedied when the analysis is done in an expanding universe.

The perturbative derivation of the Jeans instability criterion can be carried out in a cosmological setting by considering a sinusoidal perturbation superposed on a uniformly expanding background. Here, as in the Newtonian limit, there is a critical wavelength $\lambda_J$ that separates oscillatory and growing modes. Although the expansion of the background slows down the exponential growth of the amplitude to a power-law growth, the fundamental concept of a minimum mass that can collapse at any given time remains the same (see, e.g. Kolb & Turner 1990, Peebles 1993).

To be specific, let us consider an Einstein-de Sitter universe with $\Omega_{\rm dm} + \Omega_{\rm b} = 1$, where $\Omega_{\rm dm} = \bar{\rho}_{\rm dm}/\rho_{\rm c}$, $\Omega_{\rm b} = \bar{\rho}_{\rm b}/\rho_{\rm c}$, $\bar{\rho}_{\rm dm}$ is the average dark matter density, $\bar{\rho}_{\rm b}$ is the average baryonic density, and $\rho_{\rm c}$ is the critical density. We assume spatial fluctuations in the gas and dark matter densities in the form of a single spherical Fourier mode,

$$\frac{\rho_{\rm dm}(r,t) - \bar{\rho}_{\rm dm}(t)}{\bar{\rho}_{\rm dm}(t)} = \delta_{\rm dm}(t)\frac{\sin(kr)}{kr} \tag{1}$$

$$\frac{\rho_{\rm b}(r,t) - \bar{\rho}_{\rm b}(t)}{\bar{\rho}_{\rm b}(t)} = \delta_{\rm b}(t)\frac{\sin(kr)}{kr}, \tag{2}$$

where $\bar{\rho}_{\rm dm}(t)$ and $\bar{\rho}_{\rm b}(t)$ are the background densities of the dark matter and baryons, $\delta_{\rm dm}(t)$ and $\delta_{\rm b}(t)$ are the dark matter and baryon overdensity amplitudes, $r$ is the comoving radial coordinate, and $k$ is the comoving perturbation wavenumber. We assume a $\gamma$=5/3 ideal gas equation-of-state for the baryons. Initially, at time $t = t_{\rm i}$, the gas temperature is uniform $T_{\rm b}(r, t_{\rm i})$=$T_{\rm i}$, and the perturbation amplitudes are small $\delta_{\rm dm,i}, \delta_{\rm b,i} \ll 1$. We define the region inside the first zero of $\sin(kr)/(kr)$, namely $0 < kr < \pi$, as the "cloud".

The evolution of $T_{\rm b}(r,t)$ is determined by the coupling of free electrons to the Cosmic Background Radiation (CBR) through Compton scattering, by the adiabatic expansion of the gas, and by other radiative heating and cooling processes. Depending on the detailed chemical composition of the baryons, $T_{\rm b}(r,t)$ will be somewhere between the CBR temperature $T_\gamma \propto t^{-2/3}$ and the adiabatically-scaled temperature $T_{\rm ad} \propto t^{-4/3}$. In the limit of tight coupling to $T_\gamma$, the gas temperature remains uniform. On the other hand, in the adiabatic limit, the temperature develops a gradient according to the relation

$$T_{\rm b} \propto \rho_b^{(\gamma-1)}. \tag{3}$$

The evolution of $\delta_{\rm dm}(t)$ in the linear regime is described by the equation (Kolb & Turner 1990),

$$\left(\frac{d^2}{dt^2} + \frac{4}{3t}\frac{d}{dt}\right)\delta_{\rm dm} = \frac{2}{3t^2}\left(\Omega_{\rm b}\delta_{\rm b} + \Omega_{\rm dm}\delta_{\rm dm}\right) \tag{4}$$

whereas the evolution of $\delta_{\rm b}(t)$ is described by

$$\left(\frac{d^2}{dt^2}+\frac{4}{3t}\frac{d}{dt}\right)\delta_{\rm b}=\frac{2}{3t^2}\left(\Omega_{\rm b}\delta_{\rm b}+\Omega_{\rm dm}\delta_{\rm dm}\right)-\frac{k_{\rm B}T_{\rm i}}{\mu m_p}k^2\left(\frac{t_{\rm i}}{t}\right)^{(8+2\beta)/3}\left(\delta_{\rm b}+\frac{2}{3}\beta[\delta_{\rm b}-\delta_{\rm b,i}]\right). \quad (5)$$

Here, $m_p$ is the proton mass, $\mu$ is the mean molecular weight of the gas, and $k_{\rm B}$ is the Boltzmann constant. The parameter $\beta$ distinguishes between the two limits for the evolution of the gas temperature. In the adiabatic limit $\beta=1$, and when the baryon temperature is uniform and locked to the background radiation, $\beta=0$. The last term on the right hand side (in square brackets) takes into account the extra pressure gradient force in $\nabla(\rho_{\rm b}T)=(T\nabla\rho_{\rm b}+\rho_{\rm b}\nabla T)$, arising from the temperature gradient which develops according to equation (3) in the adiabatic limit. The Jeans wavelength $\lambda_J=2\pi/k_J$ is obtained by setting the right-hand side of equation (5) to zero, and solving for the critical wavenumber $k_J$. As can be seen from equation (5), the critical wavelength $\lambda_J$ (and therefore the mass $M_J$) is time-dependent [4]. We infer from equation (5) that as time proceeds, perturbations with increasingly smaller initial wavelengths stop oscillating and start to grow. To derive the time-evolution of $\lambda_J$, we must solve for the time-evolution of the temperature $T_{\rm b}(r,t)$, including the effects of the various cooling and heating mechanisms, and the evolution of the chemical composition of the baryons.

Another, more fundamental, difficulty is to relate the Jeans mass to the mass of collapsed, bound objects. The above analysis is perturbative (eqs. [4] and [5] are valid only while $\delta_{\rm b},\delta_{\rm dm}\lesssim 1$) and can only describe the initial phase of the collapse. As $\delta_{\rm b}$ and $\delta_{\rm dm}$ grow and become larger than unity, the density profiles start to evolve. In particular, since material at various radii collapses at different times, there is no reason to assume that the collapsed object includes the entire mass within the original density peak. Indeed, it has been realized (Larson 1969) that in the non-linear regime the density and temperature profiles steepen considerably and the central part of the sphere becomes exceedingly dense and hot. An object may form in the center, and subsequently accrete material from the original peak. In the limit of high masses ($\lambda\gg\lambda_{\rm Jeans}$) and in the absence of radiative cooling, the gas particles virialize through a shock that propagates outwards from the center of the sphere. The shock structure and the evolution of the density profile is self-similar, and near the center the density approaches the power-law profile $\rho\propto r^{-2.25}$ (this was first predicted by Gott 1975, and later proved via a semi-analytic approach by Bertschinger 1985). Although the initial perturbation has a definite mass, it is not clear what mass to associate with the end-product of the collapse. The relation between the mass of the bound object $M_{\rm obj}$ and the parameters of the initial perturbation can be expressed as $M_{\rm obj} = M_{\rm obj}(k, \delta_{\rm dm,i}, \delta_{\rm b,i}, t_{\rm i})$. This relation cannot be inferred directly from linear theory.

[4]In the limit $\Omega_{\rm b}\ll\Omega_{\rm dm}\approx 1$ and $T_{\rm b}\approx T_\gamma$ (i.e. $\beta=0$ in eq. [5]), the growing solution to equation (4) gives $\delta_{\rm dm}\propto t^{-2/3}$, and equation (5) can be made homogeneous in time. In this limit, i.e. when almost all the mass is in dark matter and the density of free electrons is sufficiently high to lock the gas temperature to the CBR temperature, $\lambda_J$ becomes approximately time-independent.

In the presence of radiative cooling, the dynamical evolution of the gas departs from self-similarity. In the limit of very efficient cooling, gas particles follow free-fall trajectories until they come arbitrarily close to the origin. If the cooling time is comparable to the dynamical time, the gas particles go through various phases of partial pressure support (see, e.g., Thoul & Weinberg 1995). In general, radiative cooling will facilitate the collapse of the clouds, since it reduces the pressure support of the gas. Thus we expect that $M_{\rm obj}(k, \delta_{\rm dm,i}, \delta_{\rm b,i}, t_{\rm i})$ will be increased relative to its value in the absence of cooling.

The Jeans mass derived from linear theory specifies whether an initial perturbation, characterized by the parameters $k$, $\delta_{\rm dm,i}$, $\delta_{\rm b,i}$ and $t_{\rm i}$, starts to grow. To determine the mass of the bound object resulting from its collapse we must follow numerically the nonlinear evolution of the density peak, including all microphysical gas processes. In the following sections, we address this problem using a one-dimensional spherically symmetric hydrodynamical code.

## 3. THE NUMERICAL CODE

### 3.1. *Description of the code*

The code we use is a modified version of the one-dimensional, spherically symmetric Lagrangian hydrodynamics code written by Thoul (Thoul & Weinberg 1995). The code evolves a mixture of two fluids by moving concentric spherical shells of fixed mass in the radial direction. A description of the numerical method and the details of the original code can be found in the above reference. Here we only summarize the equations solved by the code, and describe briefly the new additions: chemical reactions, Compton cooling, and cooling by the $H_2$ molecule.

The Lagrangian equations of hydrodynamics in spherical symmetry are the continuity equation for the baryons:

$$\frac{dm_{\rm b}}{dr_{\rm b}} = 4\pi r_{\rm b}^2 \rho_{\rm b}, \tag{6}$$

the momentum equations for the baryons and dark matter:

$$\frac{dv_{\rm b}}{dt} = -4\pi r_{\rm b}^2 \frac{dp}{dm_{\rm b}} - \frac{M(r_{\rm b})}{r_{\rm b}^2}, \tag{7}$$

$$\frac{dv_{\rm dm}}{dt} = -\frac{M(r_{\rm dm})}{r_{\rm dm}^2}, \tag{8}$$

the energy equation:

$$\frac{du}{dt} = -p\frac{d(\rho_{\rm b}^{-1})}{dt} + \frac{\Lambda}{\rho_{\rm b}}, \tag{9}$$

and the equation of state:

$$p = (\gamma - 1)\rho_{\rm b} u. \tag{10}$$

Here $\gamma = 5/3$ is the adiabatic index, and we have set $G = 1$. Each baryonic shell is described by its radius $r_{\rm b}$, velocity $v_{\rm b}$, mass density $\rho_{\rm b}$, pressure $p$, internal energy per unit mass $u$, and the baryonic mass interior to it $m_{\rm b}$. The dark matter shells are described by their radius $r_{\rm dm}$, velocity $v_{\rm dm}$, and the enclosed dark matter mass $m_{\rm dm}$. $M(r)$ is the total (baryonic plus dark matter) mass inside a shell at radius $r$. The details of the cooling terms included in $\Lambda$ in equation (9) are given in Appendix A.

The baryonic component is composed of the nine species H, $H^-$, $H^+$, He, $He^+$, $He^{++}$, $H_2$, $H_2^+$, and $e^-$. For the $i^{\rm th}$ species, we denote the total number of particles by $N_i$, the number density and molecular weight of these particles by $n_i$ and $\mu_i$ respectively, and the mass fraction by

$$y_i = \frac{\mu_i n_i}{\sum_{j=1}^{9} \mu_j n_j} = \frac{\mu_i n_i}{\rho_{\rm b}}. \tag{11}$$

The relative abundances of helium species ($i = 4,5,6$) are denoted by $x_i = n_i/n_{\rm He,tot}$, and the relative abundance of the hydrogen species ($i = 1,2,3,7,8$) and electrons ($i = 9$) are given by $x_i = n_i/n_{\rm H,tot}$, where $n_{\rm He,tot} = (n_{\rm He} + n_{\rm He^+} + n_{\rm He^{++}})$ and $n_{\rm H,tot} \equiv (n_{\rm H} + n_{\rm H^+} + n_{\rm H^-} + 2n_{\rm H_2} + 2n_{\rm H_2^+})$.

The above nine species interact through various chemical reactions. Only the global quantity $\rho_{\rm b}$ appears in the hydrodynamical equations (6)-(10). After each shell has been moved to its new location by a hydrodynamical iteration, the nine species are allowed to interact for a time $dt_{\rm H}$, and the relative abundances in each shell are upgraded according to the rate equations

$$\frac{dn_i}{dt} = \sum_{l=1}^{9}\sum_{m=1}^{9} a_{lmi} k_{lm} n_l n_m + \sum_{j=1}^{9} b_{ji} k_j n_j. \tag{12}$$

Here the $k_{lm}$ denote reaction rate coefficients, $k_n$ represent photo-ionization or photo-dissociation integrals, and $a_{lmi}, b_{ni} = 0, \pm 1$, or $\pm 2$, depending on the reaction. The next hydrodynamical iteration starts at a time interval $dt_{\rm H}$ later, with the updated relative abundances. In Appendix B we provide a table of the chemical reactions, and the adopted rate coefficients and cross sections.

A complication to the above numerical approach arises due to the stiffness of the system of differential equations (12). The stiffness increases the required computation time by a large factor, because standard solver routines for differential equations require exceedingly small time-steps, while routines designed to solve a stiff system of equations require many matrix operations with 9×9 matrices, in our case. We have experimented with each routine in *Numerical Recipes* (Press et al. 1992), and have found the routine *stifbs* to be the most efficient. We also compared the performance of *stifbs* with the more popular routine LSODAR (Hindmarsh 1983) to solve the system (12). We found *stifbs* to produce identical answers $\sim 20\%$ faster than LSODAR. The nine equations in system (12) could be reduced to six, using the conservation equations for the hydrogen and helium mass and for the charge. However, we found that this reduction actually *increases* the computation time, because the use of the algebraic conservation laws requires that the calculation of each number density $n_i$ be done to a much higher degree of accuracy than otherwise necessary. Another simplification could arise if the species were either "frozen", or reached equilibrium within a hydrodynamical iteration time. Unfortunately, equations (12) do not have a single time constant and the concentrations of different species change on different time-scales; in particular, $n_{\mathrm{H}_2}$ evolves on the hydrodynamic time-scale. The only simplification we were able to make was to freeze the helium abundances during the initial phase of the dynamical evolution, thus reducing the number of equations to be solved from 9 to 6.

We adopt two different central boundary conditions, corresponding to placing either a hard sphere or a sink (black hole) in the center (see Thoul and Weinberg 1995). In the first case, the dark matter virializes by bouncing off the hard sphere, while in the latter case the dark matter falls freely into a central sink. The first case may provide a more realistic representation of collapsing regions in the universe, for which non-spherical motions (or angular momentum) define an inner boundary to the infall. Quantities for the $n^{th}$ shell are always calculated using information from the $n^{th}$ and $(n+1)^{th}$ shell, thus boundary effects from the surface of the sphere propagate inwards by one shell at each iteration. In order to avoid unphysical contamination of the results by signals propagating from the outermost shell in the simulation, we choose the outer boundary condition to be such that the total perturbed mass inside the outermost shell is zero, *and* the density profile has zero slope at the surface. Both of these conditions are satisfied at the first minimum of the $\sin(x)/x$ function in equations (1)-(2), i.e. at x=4.4934. This choice eliminates both pressure and gravitational forces on the outermost shell, and allows this shell to simply follow the general expansion of the universe.

### 3.2 *Tests of the code*

Although Thoul & Weinberg (1995) have made extensive tests of the code, we performed a few additional tests. The most important side-effect of the addition of the chemical rate equations was an increase in computation time by approximately a factor of 10. This forced us to use a lower resolution (500 gas shells and 5000 dark matter shells) in our runs. However, comparisons with runs of five times higher resolution (2500 gas shells, 25000 dark matter shells) showed no considerable differences in the trajectories of the shells.

The numerical code contains three almost independent modules. These modules perform the hydrodynamical, chemical, and cooling calculations. We tested these three modules as follows.

*(a) Hydrodynamics*

The hydrodynamics part of the code was tested by Thoul & Weinberg (1995). They reproduced analytical solutions for pulsations of a polytrope, and for the self-similar collapse of a spherically symmetric, cosmological perturbation (Bertschinger 1985). In order to test the code in a context closer to the present application, we checked that it reproduces the evolution of the density contrast in the linear regime. The linear theory for the evolution of the density contrast of a single spherical Fourier mode was described in § 2 above. Supplemented with the temperature evolution $T_{\rm b}(t)$, this calculation yields the factor $\delta_k(t)$ in

$$\delta(r,t) \equiv \frac{\delta\rho}{\rho}(r,t) = \delta_k(t)\frac{\sin(kr)}{kr} \tag{13}$$

for each value of $k$ for both the dark matter and the gas. We ran the code with an initial overdensity profile given in equations (1)-(2) for various values of $k$ and $\delta_k(t=0) = 0.05$. We found that the shape of the profile stays unchanged to a high precision, while the overdensity changes by an overall factor very close to the value predicted from linear theory. Figure 1 shows how closely the overdensity in three gas shells follow the predicted $\delta_k(t)$, for four different values of $k$.

*(b) Chemistry*

To test the numerical integrations of the system of chemical reactions (12), we checked that the total hydrogen and helium mass and the total charge are conserved, or equivalently that the quantities

$$N_{\rm H,tot} = N_{\rm H} + N_{\rm H^+} + N_{\rm H^-} + 2N_{\rm H_2} + 2N_{\rm H_2^+} \tag{14}$$

$$N_{\rm He,tot} = N_{\rm He} + N_{\rm He^+} + N_{\rm He^{++}} \tag{15}$$

$$Q_{\rm tot} = N_{\rm H^+} - N_{\rm H^-} + N_{\rm He^+} + 2N_{\rm He^{++}} + N_{\rm H_2^+} - N_{\rm e}, \tag{16}$$

are constant in time in each shell. We found all three quantities to be conserved to at least one part in $10^8$. We also checked that the abundances converge to the correct equilibrium values. We ran the routine *stifbs* starting with the set of initial concentrations $x_{H_2} = 1.7 \times 10^{-6}$, $y_{He} = 0.24$, and $y_H \approx 0.76$ at various temperatures, until none of the abundances changed by more than 0.1% from one step to the next. The equilibrium abundances of $H_2, H^-$, and $H_2^+$ are always very small, so by eliminating these 3 species from equations (12), and by setting $d/dt = 0$, one can find algebraically the expected equilibrium abundances of the six remaining species. We found that *stifbs* indeed converged to these values. Figure 2 shows the equilibrium abundances of all nine species vs. temperature.

*(c) Cooling and Chemistry*

Several authors have considered molecular cooling of a uniform sphere in free-fall collapse (e.g. Palla et al. 1983, Lepp & Shull 1984). The general conclusion from these studies can be summarized as follows. On a temperature vs. density plot, the temperature departs from the adiabatic relation, $T_b \propto \rho_b^{2/3}$, due to $H_2$ cooling and achieves a local maximum $T_{max}$, after which it turns around and declines as the density increases further. The temperature does not start to rise again over several decades of increase in density. The exact values of $T_{max}$, the density at the turnaround in temperature, and the density at which the temperature starts to rise again, all depend on the adopted reaction rates and cooling function, as well as on the initial size, density, temperature and composition of the sphere.

We therefore tested only the general conclusion that a turnaround in temperature should occur and that the temperature should not increase over the next several decades in density. We considered a uniform sphere in a homogeneous free-fall collapse with the following initial parameters: $R = 10^{21}$cm, $\rho_0 = 4.52 \times 10^{-26}$g cm$^{-3}$, $T = 64$K, $x_{H_2} = 1.7 \times 10^{-6}$, $y_{He} = 0.24$, and $y_H \approx 0.76$. The chemical reactions and cooling mechanisms are summarized in Appendices A and B. The resulting temperature vs. density plot, as well as the relative abundance of $H_2$ vs. density, are shown in Figure 3. As seen in this figure, turnaround indeed occurs at $\rho/\rho_0 = 80$ when the temperature reaches the maximum value $T_{max} \sim 400$K. We note that the relative abundance of $H_2$ reaches $x_{H_2} = 3.8 \times 10^{-3}$, which is consistent with the $H_2$ fractions in Palla et al. (1983), and Lepp & Shull (1984).

## 4. NUMERICAL RESULTS AND DISCUSSION

### 4.1. Initial Conditions

For all cases presented in this paper, we adopt a Hubble constant of $H_0 = 50$ km s$^{-1}$ Mpc$^{-1}$ ($h = 0.5$). In runs with dark matter we assume an Einstein-de Sitter universe with $\Omega_{\rm dm} = 0.9$, $\Omega_{\rm b} = 0.1$, and in runs without dark matter, we assume an open universe with $\Omega_{\rm b} = 0.1$. We start the integrations at $z_i = 500$, when the perturbations are still in the linear regime. The temperature for the CBR is taken to be $T_\gamma = 2.726(1+z)$ K (Mather et al. 1994), and the initial temperature of the cloud is assumed to be uniform and equal to $T_\gamma$ at $z_i$. We use spherical $k$-modes as the initial overdensity profiles,

$$\rho_{\rm dm,i}(r) = \Omega_{\rm dm}\rho_{\rm c}[1 + \delta_{\rm dm,i}\frac{\sin(kr)}{kr}] \tag{17}$$

$$\rho_{\rm b,i}(r) = \Omega_{\rm b}\rho_{\rm c}[1 + \delta_{\rm b,i}\frac{\sin(kr)}{kr}], \tag{18}$$

where $\rho_{\rm c} = (6\pi G t^2)^{-1}$ is the critical density. With this choice, the evolution in the linear regime can be compared directly to the prediction of equations (4)-(5). As mentioned at the end of § 3.1, this choice also eliminates numerical edge effects at the outer boundary. We have experimented with alternative overdensity profiles, such as a Gaussian, or the most probable profile for random Gaussian fluctuations (Bardeen et al. 1986, hereafter BBKS), and found that our qualitative conclusions are insensitive to the detailed shape of the initial density profile.

The density profiles in equations (17)-(18) can be used to define three characteristic mass-scales. First, the total baryonic mass which is gravitationally bound includes all the baryonic material enclosed by the radius corresponding to the first *minimum* of $\sin(kr)/(kr)$, i.e. inside $kR_{\rm bound} = 4.4934$. We denote this mass as $M_{\rm bound}$. Second, we define the mass of the baryonic material which has a positive overdensity (i.e. contained within the first *zero* of $\sin(kr)/(kr)$, at $kR_{\rm cloud} = \pi$) to be $M_{\rm cloud}$. Finally, the baryonic mass which has virialized and condensed into a central object, $M_{\rm obj}$, is redshift-dependent and will be defined more precisely in the next section.

The initial amplitude of the dark matter density profile perturbation can be written as $\delta_{\rm dm,i} = \nu\sigma$, where $\sigma$ is the *r.m.s.* density fluctuation within some filter of a given shape and size in the CDM model. The value of $\nu$ characterizes the likelihood of the occurance of the peak in a Gaussian density field. This characterization depends weakly on the shape and size of the filter, and strongly on the choice of power-spectrum. For reference, we quote in Table 1 the value of $\sigma$ for six different mass-scales $M_{\rm cloud}$, and two different filter shapes with the COBE normalized BBKS power spectrum. The values in this table were obtained under the normalization $\sigma_{8h^{-1}{\rm Mpc}} = 1.22$ (for $\Omega_{\rm dm} = 0.9$, and $h = 0.5$; cf. Bunn et al. 1995), using a filter radius $R_f = 0.12 R_{\rm cloud}$. The initial amplitude of the baryon density perturbation is a small fraction of the dark matter amplitude, because the baryons had been locked to the CBR until $z \approx 10^3$, while the dark matter fluctuations have had more time to grow since matter-radiation equality. We chose the somewhat arbitrary value $\delta_{\rm b,i} = \frac{1}{10}\delta_{\rm dm,i}$ to take this effect into account. Because of the tight coupling of the baryons to the CBR before the recombination epoch, the initial velocity of each baryonic shell at $z_i = 500$ is assumed to be the Hubble velocity $v_{\rm b}(r) = Hr$. For the dark matter, we assume that the perturbation is dominated by the growing mode, and choose the initial velocity for the dark matter shells accordingly, $v_{\rm dm}(r) = Hr(1 - \frac{1}{3}\langle\delta_{\rm dm}\rangle_r)$.

The initial relative abundances and mass fractions of the various species are taken to be: $x_{\rm H^+}$=$x_{\rm e}$=$10^{-4}$ (cf. Peebles 1993), $x_{\rm H_2}$=$1.7 \times 10^{-6}$ (cf. Palla et al. 1984), $x_{\rm H^-}$=$x_{\rm He^+}$=$x_{\rm He^{++}}$=$x_{\rm H_2^+}$=0, and $(y_{\rm H}+y_{\rm H^+}+y_{\rm H_2}+y_{\rm e})$=0.76, $y_{\rm He}$=0.24.

### 4.2. Definitions of "Collapse" and "Object"

In order to express the state of the gas as a function of initial conditions, we need to define what we mean by the concepts of "collapse" and "object". We base our heuristic definitions on the spherical top-hat collapse model. In this model, virialization occurs at half the turnaround radius when the overdensity is $\sim 100 - 200$. The value appears to be rather insensitive to the underlying cosmology (cf. Fig. 4 in Richstone, Loeb & Turner 1992). We therefore define a shell to have *collapsed* if the average overdensity enclosed by that shell is $\bar{\delta}_{\rm b} \geq \bar{\delta}_{\rm thr} = 200$.

As time evolves, the density profile within the collapsing region steepens, and at some definite time, the average density contrast within the innermost baryonic shell reaches the threshold value $\bar{\delta}_{\rm b}(r_1) = \bar{\delta}_{\rm thr} = 200$. The subsequent shells reach this threshold some time later. We define an *object* to include all the baryonic material within the outermost shell that has collapsed. According to this definition, the mass of the object $M_{\rm obj}(z)$ is zero until the first shell reaches $\bar{\delta}_{\rm thr} = 200$, and is monotonically increasing aftwerwards.

Unlike definitions based on optical depth, or absolute density, the threshold overdensity does not strongly depend on the mass-scale involved. In the presence of cooling we find our results to be rather insensitive to the exact value of the threshold overdensity (see below).

### 4.3. Collapse without dark matter or cooling

The simplest system to consider consists of a purely baryonic cloud following adiabatic infall, without any dark matter or radiative processes. We present results from test cases of such a system in order to establish a prototype against which we will compare results later, when the dark matter component and the radiative processes will be added.

For the initial conditions (cf. § 4.1), it is necessary to fix two free parameters: the height of the density peak $\delta_{\rm b,i}$, and the mass-scale $M_{\rm cloud}$ or equivalently the width $\sim k^{-1}$ of the initial density peak. $\delta_{\rm b,i}$ sets the redshift at which the perturbation reaches the nonlinear phase, and $k$ determines the initial ratio of pressure to gravitational forces.

In Table 2 we summarize the parameters for eight different runs. The runs numbered 1-4 are purely baryonic and do not include any radiative cooling, while the runs numbered 5-8 contain a mixture of baryons and cold dark matter, and include radiative cooling.

We first discuss the results of the first four simulations. In Figure 4 we show the trajectories of 10 shells (enclosing 7, 17, 27, ..., 87, and 97% of the total baryonic mass $M_{\rm bound}$) across the radius of the sphere (solid curves), along with the trajectories that each shell would have had if it were freely falling (dashed curves). In all cases, the baryonic shells expand, turn around, collapse, shock, and then drift slowly towards the center. The trajectories of the shells are similar to the semi-analytic solution of Bertschinger (1985). Here, however, the shells hit the shock at slightly different fractions of their turnaround radii rather than at the fixed fraction of $\frac{1}{2}$, due to corrections of order $\delta_i^2$ to Bertschinger's results. The higher mass runs (numbers 1 and 3) initially follow the free-fall trajectories, but eventually deviate from them after the shells shock. In the lower mass runs (numbers 2 and 4) pressure is important and the collapse is delayed relative to the free-fall case even before the shock. The temperature evolution for these shells is shown in Figure 5.

In Figure 6 we show the fraction of the total bound baryonic mass that has collapsed to form the object (solid lines). We also show what this collapsed fraction would be for a pressureless freely-falling gas (dotted lines). The higher mass runs deviate from the free-fall curves only after the shocked shells collapse. In the lower mass runs, the collapse is delayed relative to the free-fall case at all times.

### 4.4. Collapse with dark matter and cooling

Given some choice of cosmological parameters $H_0$, $\Omega_{\rm dm}$ and $\Omega_{\rm b}$, and the ratio $\delta_{\rm b,i}/\delta_{\rm dm,i}$, there are again only two free parameters left, namely the initial baryonic cloud mass $M_{\rm cloud}$, and the amplitude $\delta_{\rm dm,i}$. We have performed simulations on a grid in the $(M_{\rm cloud}, \delta_{\rm dm,i})$ parameter space. The details of the four simulations numbered 5-8 in Table 2 are shown in Figures 7-12.

Figure 7 shows the trajectories of 10 shells (enclosing 7, 17, 27, ..., 87, and 97% of the total baryonic mass $M_{\rm bound}$). The trajectory of each shell is shown by a solid line, until the average density contrast within it is smaller than the threshold value $\bar{\delta}_{\rm thr} = 200$. Once the overdensity within a shell reaches $\bar{\delta}_{\rm thr}$, it is considered to be part of the central object. To indicate this in the plots, we continue to draw the shell trajectories as dotted lines after they reach $\bar{\delta} = \bar{\delta}_{\rm thr}$. The dashed lines show the trajectories of the dark matter shells. In the smallest mass run (number 5) cooling is inefficient and the gas shells stop moving after they shock. In larger mass cases (runs 6 and 7) cooling is more efficient. Finally, in the largest mass case (run 8) all the shocked shells cool rapidly and fall towards the center.

Figure 8 shows the evolution of the temperature with redshift for the same gas shells. The dotted lines show the CBR temperature $T_\gamma \propto (1+z)$, and the dashed lines show the adiabatic temperature scaling in the linear regime $T_{\rm ad} \propto (1+z)^2$. The temperature of the gas remains approximately uniform initially, with a value in-between the adiabatic and CBR cases. After turnaround the temperature increases adiabatically and a temperature gradient develops. Eventually, as each shell encounters the shock, its temperature jumps abruptly to the virial temperature. In the absence of cooling, the temperature of each shell remains approximately constant, increasing slowly due to the slight adiabatic compression. When cooling is efficient, i.e., when the cooling time becomes smaller than the dynamical time, the temperature reaches a maximum value and then decreases very quickly. The temperature at which this turnover occurs in our simulations is of the order $\log(T/{\rm K}) \approx 2.3 - 2.8$ at the onset of $H_2$ cooling. Figure 9 shows the evolution of the four most efficient cooling and heating terms with redshift in a typical run. The most efficient radiative mechanism before the shock (when the temperature of the gas is below the temperature of the CBR) is Compton heating; this process keeps the gas temperature above $T_{\rm ad}$. After the shock, the most efficient radiative mechanism is $H_2$ cooling, causing a turnover in the temperature. Note that the gas temperature reaches a value above $T_\gamma$, so that the $H_2$ molecules as well as Compton scattering cool the gas, rather than heat it. After the temperature starts to decline, the shells fall towards the center essentially at the free-fall rate. When the shells reach a very small radius and the numerical timestep necessary to resolve their motion becomes too small, we artificially set the temperature of these shells to $T = 200$ K and subsequently ignore them.

Figure 10 shows the evolution of the relative abundances of $H_2$ and $e^-$ within five different shells (enclosing 2, 7, 12, 17, and 22% of the total baryonic mass $M_{\rm bound}$). The highest relative abundance of $H_2$ is achieved in the case of the largest perturbed mass (run 8), where it reaches about $10^{-3}$. The abundances develop a gradient across the sphere, which is positive for $H_2$ and negative for the electrons. The evolution of the abundances are relatively universal and look similar in all four runs. In particular, $\log x_{H_2} \approx -3.6$ in all four runs before the gradient develops, and the central value is $\log x_{\rm H_2} \approx -3.2$ afterwards. The main difference between the four runs is the redshift at which the gradients develop.

Figure 11 shows typical density, temperature, and relative $H_2$ and $e^-$ abundance profiles across the sphere at redshift $z = 10$ for each of the four runs. The $x_{\rm H_2}$ profiles are almost flat at the outer part of the sphere, and well approximated by a power law $x_{\rm H_2} \propto r^{-0.4}$ within the post-shock region, while the $e^-$ fraction profiles behave less regularly. Note that the power-law profile for the density $\rho \propto r^{-2.25}$ predicted by the self-similar solution (Bertschinger 1985) is obtained near the center (see § 2 and the thick line in panel 11a).

Figure 12 shows the fraction of the collapsed baryonic mass out of the total bound baryonic mass vs. redshift for the runs 5-8, for two different values of the overdensity threshold for forming objects $\bar{\delta}_{\rm thr}$. The collapsed baryonic mass versus redshift curves are rather insensitive to the value of $\bar{\delta}_{\rm thr}$, especially for the objects that collapse early.

### 4.5. Collapse onto a black hole

Next, we perform the same runs as described in §4.4, but with a different central boundary condition; we replace the reflecting hard sphere with a black hole sink of a fixed radius. We choose the radius of the sink to be a small fraction of the initial radius of the first shell. The shells which cross the black hole radius are assigned arbitrary values of density and temperature, and are subsequently ignored. Under these conditions, the dark matter component does not bounce back and virialize since it is absorbed by the central black hole. Due to the post-shock shell crossing, the gas component continues to collapse after it shocks, unlike in the case of a reflecting sphere. This leads to a continuous rise in the gas temperature after the shock, enabling more efficient post-shock cooling.

In the presence of a central black hole, we have found an oscillatory instability of the shock front in the cases involving efficient radiative cooling. Instead of propagating outwards monotonously as in the cases shown in Figure 7, the shock front in these cases can reverse its motion and oscillate. However, for the present application, the details of these oscillations can be glossed over, as we are only interested in the question whether shells reach $\bar{\delta} = \bar{\delta}_{\rm thr}$. The oscillation of the shock front occurs only for shells that have already collapsed according to our definition, hence this instability affects only the structural details of the objects we find but not the question of whether they form.

### 4.6. Discussion

We next explore the $(M_{\rm cloud}, \delta_{\rm dm,i})$ parameter space for the two different boundary conditions. In each case, we determine whether radiative cooling affects the dynamics of the gas and whether an object forms.

To assess the significance of cooling in each run, we compared the trajectories of the gas shells with and without cooling. If the trajectories in the two cases were indistinguishable, we concluded that cooling had no effect; if they differed significantly, we concluded that cooling was efficient. To establish whether or not an object formed, we used a simple criterion: for an object to form, we required that at least half of the baryonic mass within the original cloud should collapse by a redshift $z = 10$, i.e., $M_{\rm obj}(z = 10)/M_{\rm cloud} \geq \frac{1}{2}$. This corresponds to 1/6 of the total bound mass ($M_{\rm bound}$) being above a density contrast of 200.

The main parameters which determine the post-shock cooling rate are the virial temperature $T_{\rm vir}$, and the abundance of molecular hydrogen $x_{\rm H_2}$ after the shock. While it is not possible to calculate $x_{\rm H_2}$ analytically, the scaling of the virial temperature can be derived from the spherical top-hat model (White 1994). Using our runs with a central hard sphere to calibrate this relation for the baryons, we find that for $\Omega_0 = \Omega_{\rm b} + \Omega_{\rm dm} = 1$, the post-shock temperature of the gas shells closely follow

$$T_{\rm vir} = 55(1 + z_s) \left(\frac{h}{0.5}\right)^{2/3} \left(\frac{M_{\rm cloud}}{10^5 M_\odot}\right)^{2/3} \ {\rm K}, \qquad (19)$$

where $z_s$ is the shock redshift of each gas shell, and $M_{\rm cloud}$ is the baryonic mass of the cloud. The post-shock values for $x_{\rm H_2}$ do not vary substantially between runs, as shown in Figure 10.

In Figure 13 we show the grid of points we explored in the $(M_{\rm cloud}, \delta_{\rm dm,i})$ parameter space. The solid lines correspond to lines of constant central virial temperatures, $T_{\rm vir} = 85$ K and $T_{\rm vir} = 120$ K. The solid circles on this grid represent cases where cooling had no effect; the open circles represent cases where cooling was very efficient; and the open triangles represent intermediate cases, where cooling had a small but noticeable effect. The $(M_{\rm cloud}, \delta_{\rm dm,i})$ plane can be divided into two regions: the region where cooling is important, and the region where cooling is not important. According to our results, this division follows roughly a line of constant virial temperature $T_{\rm vir} \sim 10^2$ K.

In Figure 14 we show again the above grid of points, but here the solid dots represent cases where no object formed by $z = 10$; the open dots represent cases where objects formed by $z = 10$; and the triangles represent intermediate cases where objects are just starting to form at $z = 10$. Again, we can divide the $(M_{\rm cloud}, \delta_{\rm dm,i})$ plane into two regions, depending on whether objects form or not. The constant $T_{\rm vir}$ curves from Figure 13 are also overlayed on this plane. According to the upper panel, we see that the division based on cooling goes through points of somewhat higher $\delta_{\rm dm,i}$ than the division based on collapse in the case of a central reflecting sphere. In other words, the transition from "no collapse" to "collapse" takes place in a regime where cooling is ineffective. Thus, radiative cooling does not introduce new mass or overdensity scales that are able to collapse.

The above result can be understood physically as follows. In order for $H_2$ cooling to affect the dynamics, the gas temperature $T_{\rm b}$ must satisfy two conditions. First, in order for cooling (rather than heating) to occur, the gas temperature must be higher than the CBR temperature. Secondly, the $H_2$ cooling time needs to be smaller than the typical dynamical time; we find that this is only possible for a temperature $T_{\rm b} \gtrsim 10^2$ K. We find that the above two conditions cannot be satisfied simultaneously before the gas shocks. The gas temperature drops below $10^2$K as early as $z \sim 100$, before any object forms, and rises above $10^2$K again only after the gas shocks (cf. Fig. 8). However, by that late time the gas has already achieved an overdensity close to our collapse criterion $\bar{\delta} \sim 200$. Thus, even if cooling alters the post-shock trajectories significantly (as in run 8 in Fig. 7), its effect translates into a minor change in the collapsed baryonic mass as a function of redshift, $M_{\rm obj}(z)$.

The situation is somewhat different for collapses with a central sink (black hole) boundary condition. In the hard sphere case, the dark matter shells bounce back from the center, virialize, and stop moving subsequently. After a gas shell shocks, it is not crossed by dark matter-shells, unless radiative cooling is sufficiently effective to reduce its pressure support and allow it to fall towards the center. In the central sink case, the dark matter shells do not bounce back and virialize; instead, they are absorbed by the sink. Even after a gas shell shocks, it is crossed continuously by new dark matter shells. Therefore, gas shells continue to collapse after they shock even in the absence of radiative cooling. The continued infall heats the gas adiabatically, and the importance of post-shock cooling is enhanced relative to the hard-sphere case. The lower panel of Figure 13 shows that cooling is indeed effective for much smaller mass scales than in the hard-sphere case; the gas dynamics in clouds with a baryonic mass as low as $M_{\rm cloud} = 10^{2-3} M_\odot$ is altered by cooling. For higher mass runs, $H_2$ cooling becomes so effective as to smooth-out the shocks; the gas shells fall-in rapidly and their trajectories show little evidence for a discontinuity. The cooling effects change the collapsed baryonic mass as a function of redshift, $M_{\rm obj}(z)$, and allow slightly lower mass-scales to collapse by $z = 10$. Indeed, the lower panel of Figure 14 shows that the transition from "no collapse" to "collapse" occurs for slightly lower masses than in the hard-sphere case, and in a regime where cooling is efficient.

## 5. CONCLUSIONS

In this paper we have investigated the early ($z \gtrsim 10$) formation of bound objects with masses in baryons comparable to the linear-theory Jeans-mass ($\sim 10^5 M_\odot$). We used a spherically symmetric Lagrangian code to analyze the dynamics of low-mass overdense regions in a cold dark matter cosmology. The code solved for the radial trajectories of the dark matter and gas shells, along with the density, temperature, and non-equilibrium abundances of various chemical ingredients of the gas in each shell.

The dynamics of a primordial gas cloud depends primarily on two parameters, the baryonic mass ($M_{\rm cloud}$) and the initial overdensity ($\delta_{\rm dm,i}$) of the cloud. We have computed the evolution of gas clouds on a grid of 48 runs, corresponding to different points in the ($M_{\rm cloud}$, $\delta_{\rm dm,i}$) parameter-space. We have done the calculations with two different types of central boundary conditions for the dark matter: a reflecting hard sphere, and a central sink (black hole).

Based on the above grid of runs, one can study the effect of radiative cooling by molecular hydrogen ($H_2$) on the dynamics of the baryons. We find that, in the hard sphere case, $H_2$ cooling is important only when the virial temperature of the collapsing object is $\gtrsim 100$ K (cf. eq. [19] and Fig. 13). However, this temperature regime is accessible only after shock-heating, when the gas has already condensed to be part of a nonlinear object. Before the gas enters the virialization shock, its temperature is lower than the CBR temperature and radiative $H_2$ cooling is not possible. This implies that contrary to previous suggestions, radiative cooling could not have triggered the formation of the first generation of bound objects in the universe.

However, the evolution of the gas after it shocks can be substantially altered due to $H_2$ cooling. This effect is particularly emphasized in the case of a central-sink boundary condition (cf. lower panel of Fig. 13), where the dark matter shells cross the baryonic shells and pull them effectively towards the center. This process brings the baryonic shells to a thermodynamic state that favors efficient cooling.

In general, we find that the presence of shell crossing by dark matter allows baryonic objects with masses well below the linear-theory Jeans mass to form. Figures 7 and 14 demonstrate that the cores of objects containing $\lesssim 10^3 M_\odot$ in baryons can collapse in a standard CDM cosmology by a redshift $z = 10$. It is therefore possible that high-mass stars had formed in the intergalactic space directly as a result of the primordial cosmological perturbations, and not just due to the commonly discussed process of fragmentation inside more massive objects. Intergalactic stars may have left their signature on the metalicity of the intergalactic medium and of Ly$\alpha$ absorption clouds through supernovae at high redshifts. It is difficult to detect the light from intergalactic stars unless they are assembled into high surface-brightness objects. However, such stars should result in gravitational microlensing events of quasars, with a lensing probability comparable to their cosmological density parameter $\sim \Omega_\star$ (Press & Gunn 1973).

The most significant limitation of our approach is the assumption of spherical symmetry. This assumption is forced upon us by practical computational considerations. Spherical symmetry may, in fact, apply to the collapse of a small subset of systems in environments with unusually low shear (Eisenstein & Loeb 1995b). However, most three-dimensional collapses are likely to be triaxial (Eisenstein & Loeb 1995a, and references therein). The early pancaking along the short axis would enhance the gas density and the corresponding $H_2$ formation, and could lead to a more efficient cooling of the gas. Fragmentation may lead to similar effects (see Ruzmaikina 1973, and Kashlinsky & Rees 1983). Future three-dimensional simulations are essential in exploring this more complicated dynamics.

We thank Daniel Eisenstein, Ignasi Forcada I Miro, Ofer Lahav, Steve Lepp, and Martin Rees for valuable discussions, and Alex Dalgarno and Tom Abel for their detailed comments on the chemical reactions. ZH acknowledges financial support from a Cambridge European Trust ORS award and the Isaac Newton Studentship. AT was supported by the Ambrose Monell Foundation and by NSF grant PHY92-45317.

## APPENDIX A: COOLING AND HEATING TERMS

The cooling terms in $\Lambda$ of equation (9) in § 3.1 are given by the sum of the following 12 terms below, in units of ergs $s^{-1}$ $cm^{-3}$. References are Black (1981) and Cen (1992) for the first 10 terms, Spitzer & Hart (1971) for the Gaunt factor $g_{\rm ff}$ appearing in $\Lambda_{\rm Brem}(T)$, Shapiro & Kang (1987) for $\Lambda_{\rm Comp}(T)$, and Lepp & Shull (1983) for $\Lambda_{\rm H_2}(T)$. In these formulas, $T$ denotes the gas temperature in K, and $T_\gamma$ the temperature of the CBR, $T_\gamma = 2.726(1+z)$ K. The number densities, $n_i$, are in $cm^{-3}$.

*Collisional ionization of* H, He, *and* $He^+$*:*

$$\Lambda_{\rm H}(T) = 1.27\times10^{-21}T^{\frac{1}{2}}(1+T_5^{\frac{1}{2}})^{-1}\exp(-157809.1/T)n_{\rm e}n_{\rm H}$$
$$\Lambda_{\rm He}(T) = 9.38\times10^{-22}T^{\frac{1}{2}}(1+T_5^{\frac{1}{2}})^{-1}\exp(-285335.4/T)n_{\rm e}n_{\rm He}$$
$$\Lambda_{\rm He^+}(T) = 4.95\times10^{-22}T^{\frac{1}{2}}(1+T_5^{\frac{1}{2}})^{-1}\exp(-631515/T)n_{\rm e}n_{\rm He^+}$$

*Recombination to* H, He, *and* $He^+$*:*

$$\Lambda_{\rm H^+}(T) = 8.70\times10^{-27}T^{\frac{1}{2}}T_3^{-0.2}(1+T_6^{0.7})^{-1}n_{\rm e}n_{\rm H^+}$$
$$\Lambda_{\rm He^+}(T) = 1.55\times10^{-26}T^{0.3647}n_{\rm e}n_{\rm He^+}$$
$$\Lambda_{\rm He^{++}}(T) = 3.48\times10^{-26}T^{\frac{1}{2}}T_3^{-0.2}(1+T_6^{0.7})^{-1}n_{\rm e}n_{\rm He^{++}}$$

*Dielectric Recombination to* He*:*

$$\Lambda_{\rm He^+}(T) = 1.24\times10^{-13}T^{-1.5}\exp(-470000/T)(1+0.3\exp(-94000/T))n_{\rm e}n_{\rm He^+}$$

*Collisional excitation of* H, *and* $He^+$*:*

$$\Lambda_{\rm H}(T) = 7.50\times10^{-19}(1+T_5^{\frac{1}{2}})^{-1}\exp(-118348/T)n_{\rm e}n_{\rm H}$$
$$\Lambda_{\rm He^+}(T) = 5.54\times10^{-17}T^{-0.397}(1+T_5^{\frac{1}{2}})^{-1}\exp(-473638/T)n_{\rm e}n_{\rm He^+}$$

*Bremsstrahlung:*

$$\Lambda_{\rm Brem}(T) = 1.42\times10^{-27}g_{\rm ff}T^{\frac{1}{2}}n_{\rm e}(n_{\rm H^+}+n_{\rm He^+}+4n_{\rm He^{++}})$$
$$g_{\rm ff} = 1.10+0.34\exp(-(5.50-\log T)^2/3.0)$$

*Compton cooling:*

$$\Lambda_{\rm Comp}(T) = -1.017\times10^{-37}T_\gamma^4(T-T_\gamma)n_{\rm e}$$

*Cooling due to* $H_2$ *molecules:*

A lengthy expression for the term $\Lambda_{\rm H_2}(T)$ was adopted from Lepp & Shull (1983). The expression in this reference is an analytical fit to the authors' numerical results. It includes cooling from the following mechanisms involving $H_2$:

- dissociation of $H_2$ molecules from $H_2 - H_2$ collisions,
- dissociation of $H_2$ molecules from $H_2 - H$ collisions,
- de-excitation from excited rotational and vibrational states, due to $H_2 - H_2$ collisions,
- de-excitation from excited rotational and vibrational states, due to $H_2 - H$ collisions.

However, radiative cooling by $H_2$ is not possible thermodynamically when $T < T_\gamma$; instead $H_2$ is a source of heating under this condition. In order to correct for the presence of radiative heating from the inverse of the above processes, we multiply the quoted value $\Lambda_{H_2}(T)$ by the factor $(T - T_\gamma)/(T + T_\gamma)$. Note that there is a typographical error in Lepp & Shull 1983 (confirmed by S. Lepp, private communication), which changes the value of $\Lambda_{H_2}(T)$ considerably. In their equation (8) on p. 581, $1.10 \times 10^{-13}$ should be replaced by $1.10 \times 10^{-18}$.

## APPENDIX B: REACTION RATES AND CROSS SECTIONS

| | Reaction | Rate Coefficient ($cm^3$ $sec^{-1}$) | Reference |
|---|---|---|---|
| (1) | $H + e^- \rightarrow H^+ + 2e^-$ | $5.85 \times 10^{-11} T^{\frac{1}{2}} \exp(-157809.1/T)(1 + T_5^{\frac{1}{2}})^{-1}$ | [2] |
| (2) | $He + e^- \rightarrow He^+ + 2e^-$ | $2.38 \times 10^{-11} T^{\frac{1}{2}} \exp(-285335.4/T)(1 + T_5^{\frac{1}{2}})^{-1}$ | [1] |
| (3) | $He^+ + e^- \rightarrow He^{++} + 2e^-$ | $5.68 \times 10^{-12} T^{\frac{1}{2}} \exp(-631515.0/T)(1 + T_5^{\frac{1}{2}})^{-1}$ | [2] |
| (4) | $H^+ + e^- \rightarrow H + h\nu$ | $8.40 \times 10^{-11} T^{-\frac{1}{2}} T_3^{-0.2} (1 + T_6^{0.7})^{-1}$ | [2] |
| (5) | $He^+ + e^- \rightarrow He + h\nu$ | see expression in reference | [2] |
| (6) | $He^{++} + e^- \rightarrow He^+ + h\nu$ | $3.36 \times 10^{-10} T^{-\frac{1}{2}} T_3^{-0.2} (1 + T_6^{0.7})^{-1}$ | [2] |
| (7) | $H + H^+ \rightarrow H_2^+ + h\nu$ | see expression in reference | [10] |
| (8) | $H_2^+ + H \rightarrow H_2 + H^+$ | $6.40 \times 10^{-10}$ | [7] |
| (9) | $H + e^- \rightarrow H^- + h\nu$ | $5.57 \times 10^{-17} T^{\frac{1}{2}}$ | [9] |
| (10) | $H + H^- \rightarrow H_2 + e^-$ | $1.30 \times 10^{-9}$ | [4] |
| (11) | $H_2^+ + e^- \rightarrow 2H$ | $1.68 \times 10^{-8} (T/300)^{-0.29}$ | [8] |
| (12) | $H_2^+ + H^- \rightarrow H_2 + H$ | $5.00 \times 10^{-6} T^{-\frac{1}{2}}$ | [3] |
| (13) | $H^- + H^+ \rightarrow 2H$ | $7.00 \times 10^{-7} T^{-\frac{1}{2}}$ | [3] |
| (14) | $H_2 + e^- \rightarrow H + H^-$ | $2.70 \times 10^{-8} T^{-\frac{3}{2}} \exp(-43000/T)$ | [5] |
| (15) | $H_2 + H \rightarrow 3H$ | see expression in reference | [11] |
| (16) | $H_2 + H_2 \rightarrow H_2 + 2H$ | see expression in reference | [11] |
| (17) | $H_2 + H^+ \rightarrow H_2^+ + H$ | $2.4 \times 10^{-9} \exp(-21200/T)$ | [13] |
| (18) | $H_2 + e^- \rightarrow 2H + e^-$ | $4.38 \times 10^{-10} \exp(-102000/T) T^{0.35}$ | [13] |
| (19) | $H^- + e^- \rightarrow H + 2e^-$ | $4.00 \times 10^{-12} T \exp(-8750/T)$ | [13] |
| (20) | $H^- + H \rightarrow 2H + e^-$ | $5.30 \times 10^{-20} T^{2.17} \exp(-8750/T)$ | [13] |
| (21) | $H^- + H^+ \rightarrow H_2^+ + e^-$ | see expression in reference | [13] |
| | | Cross-section ($cm^2$) | |
| (22) | $H + h\nu \rightarrow H^+ + e^-$ | $\sigma(\nu) = a[\beta(\frac{\nu}{\nu_T})^{-s} + (1-\beta)(\frac{\nu}{\nu_T})^{-s-1}]$ $\nu \geq \nu_T$ | [12] |
| (23) | $He + h\nu \rightarrow He^+ + e^-$ | " | [12] |
| (24) | $He^+ + h\nu \rightarrow He^{++} + e^-$ | " | [12] |
| (25) | $H^- + h\nu \rightarrow H + e^-$ | " | [12] |
| (26) | $H_2^+ + h\nu \rightarrow H + H^+$ | $\sigma(\nu, T) = (p_1 + p_2\nu) \exp((\nu - p_3)^2 / 2p_4^2)$ | [14] |
| (27) | $H_2 + h\nu \rightarrow H_2^+ + e^-$ | see expression in reference | [13] |
| (28) | $H_2 + h\nu \rightarrow 2H$ | $\sigma(\nu) = 3 \times 10^{-22}$ for 11.26 eV $< h\nu <$ 13.6 eV | [6] |

1. Black (1978); 2. Cen (1992); 3. Dalgarno & Lepp (1987); 4. de Jong (1972); 5. Hirasawa (1969); 6. Hollenbach & McKee (1979); 7. Karpas et al. (1979); 8. Nakashima et al. (1987); 9. Rawlings (1988); 10. Rawlings (1993); 11. Lepp & Shull (1983); 12. Osterbrock (1974); 13. Shapiro & Kang (1987); 14. Stancil (1994).

The table above lists the chemical reactions included in equations (12) in § 3.1. We attempted to include all possibly important reactions. The rates for reactions (5), (7), (15), (16) and (21) and the cross-section for reaction (27) are given by more lengthy expressions, which can be found in the references. The expression for the cross-section in reaction (26) is a fit to the data from Stancil (1994), with temperature-dependent coefficients $p_i$. The coefficients $\alpha$, $\beta$, $\nu_T$, and $s$ appearing in the cross-sections for reactions (22)-(25) can be found in Osterbrock (1974).

Table 1: The *r.m.s.* fluctuation $\sigma(M)$ in two different shaped filters with the COBE-normalized ($\sigma_8 = 1.22$) BBKS power-spectrum of a cold dark matter universe at $z$=500.

| $M_{\rm cloud}$ | $\sigma_{\rm dm}(M, z = 500)$ | |
|---|---|---|
| $(M_\odot)$ | Top-hat | Gaussian |
| 50 | 0.118 | 0.113 |
| 100 | 0.115 | 0.110 |
| 500 | 0.107 | 0.103 |
| 1,000 | 0.104 | 0.100 |
| 5,000 | 0.097 | 0.093 |
| $10^4$ | 0.093 | 0.090 |

Table 2: Parameters of the numerical runs. Initial heights and widths are at $z = 500$.

| Run Number | Baryon Mass $M_{\rm cloud}(M_\odot)$ | Height $\delta_{\rm dm,i}$ | Width $R_{\rm cloud}$(kpc) | Baryons $\Omega_{\rm b}$ | Dark Matter $\Omega_{\rm dm}$ | Cooling Yes/No |
|---|---|---|---|---|---|---|
| 1 | $5 \times 10^7$ | 0.13 | 235.56 | 0.1 | 0 | No |
| 2 | $5 \times 10^5$ | 0.13 | 50.75 | 0.1 | 0 | No |
| 3 | $5 \times 10^7$ | 0.39 | 228.96 | 0.1 | 0 | No |
| 4 | $5 \times 10^5$ | 0.39 | 49.33 | 0.1 | 0 | No |
| 5 | 500 | 0.26 | 5.13 | 0.1 | 0.9 | Yes |
| 6 | $1,000$ | 0.26 | 6.46 | 0.1 | 0.9 | Yes |
| 7 | $1,000$ | 0.39 | 6.45 | 0.1 | 0.9 | Yes |
| 8 | $5,000$ | 0.26 | 11.05 | 0.1 | 0.9 | Yes |

## REFERENCES

Bardeen, J.M., Bond, J.R., Kaiser, N., & Szalay, A.S. 1986, ApJ, 304, 15 (BBKS)

Bertschinger, E. 1985, ApjS, 58, 39

Binney, J., & Tremaine, S. 1987, Galactic Dynamics, (Princeton: Princeton Univ. Press), pp. 287-291

Black, J.H. 1978, ApJ, 222, 125

Black, J.H. 1981, MNRAS, 197, 553

Bunn, E. F., Scott, D., & White, M. 1995, ApJ, 441, L9

Cen, R. 1992, ApJ, 78, 341

Cen, R., & Ostriker, J.P. 1994, ApJ, 431, 451

de Jong, T. 1972, A&A, 20, 263

Dalgarno, A., & Lepp, S. 1987 in Astrochemistry, ed. M. S. Varya & S. P. Tarafdar (Dordrecht:Reidel), 109

Efstathiou, G., Bond, J.R., & White, S.D.M. 1992, MNRAS, 258, 1p

Eisenstein, D.J., & Loeb, A. 1995a, ApJ, 439, 520

Eisenstein, D.J., & Loeb, A. 1995b, ApJ, 443, 11

Fan, X.M. 1994, PhD Thesis, Columbia Univ., New York

Gott, J.R. 1975, ApJ, 201, 296

Hernquist, L., Katz, N., & Weinberg, D. 1995, ApJ, 442, 57

Hindmarsh, A.C. 1983, in Scientific Computing, ed. R.S. Stepleman et al. , (North-Holland:Amsterdam), pp. 55-64.

Hirasawa, T. 1969, Prog. Theor. Phys., v. 42, no. 3, p. 523

Hollenbach, D., & McKee, C.F. 1979, ApJS, 41, 555

Jeans, J. H. 1928, Astronomy and Cosmogony (Cambridge University Press: Cambridge, UK)

Karpas, Z., Anicich, V., and Huntress, W.T. 1979, J. Chem. Phys., 70, 2877

Kashlinsky, A., & Rees, M.J. 1983, MNRAS, 205, 955

Kolb, E.W., & Turner, M.S. 1990, The Early Universe (Addison-Wesley: Redwood City, CA), pp. 344-348

Lahav, O. 1986, MNRAS, 220, 259

Larson, R.B. 1969, MNRAS, 145, 271

Lepp, S., & Shull, J.M. 1983, ApJ, 270, 578

Lepp, S., & Shull, J.M. 1984, ApJ, 280, 465

Mather, J.C., et al. 1994, ApJ, 420, 439

Matsuda, T., Sato, H., & Takeda, H. 1969, Prog. Theor. Phys., v. 42, no. 2, p. 219

Nakashima, K., Takayi, H., Nakamura, H. 1987, J. Chem. Phys., 86, 726

Osterbrock, D.E. 1974, Astrophysics of Gaseous Nebulae (Freeman and Co: San Francisco, CA)

Palla, F., Salpeter, E.E., & Stahler, S.W. 1983, ApJ, 271, 632

Press, W.H., & Gunn, J.E. 1973, ApJ, 185, 397

Press, W.H., Teukolsky, S.A., Vetterling, W.T., & Flannery, B.P. 1992, Numerical Recipes in Fortran, 2nd ed., (Cambridge University Press: Cambridge, UK)

Peebles, P.J.E. 1993, Principles of Physical Cosmology (Princeton University Press: Princeton) pp. 179, 635-636

Peebles, P.J.E., & Dicke, R.H. 1968, ApJ, 154, 891

Rawlings, J.M.C. 1988, MNRAS, 232, 507

Rawlings, J.M.C., Drew, J.E., Barlow, M.J. 1993, MNRAS, 265, 968

Richstone, D., Loeb, A., & Turner, E.L. 1992, ApJ, 393, 477

Ruzmaikina, T.V. 1973, Sov. Astron., v. 16, no. 6, p. 991

Saslaw, W.C., & Zipoy, D. 1967, Nature, 216, 976

Shapiro, P.R., & Kang, H. 1987, ApJ, 318, 32
Spitzer, L. Jr., & Hart, M.H. 1971, ApJ, 166, 483
Stancil, P.C. 1994, Harvard-Smithsonian (CfA) Preprint No. 3821
Thoul, A.A., & Weinberg, D.H. 1995, ApJ, 442, 480
Tytler, D. & Fan, X.M. 1995, ApJ, in press
White, S.D.M. 1994, Les Houches Lectures, in press, MPA preprint 831
White, M., Scott, D., & Silk, J. 1994, ARA& A, 32, 319
Woltjer, L. 1990, in Active Galactic Nuclei, SAAS-FEE Adv. Course 20, Swiss Society for Astroph. & Astron., ed. T. J. L. Courvoisier & M. Mayor (Berlin: Springer)

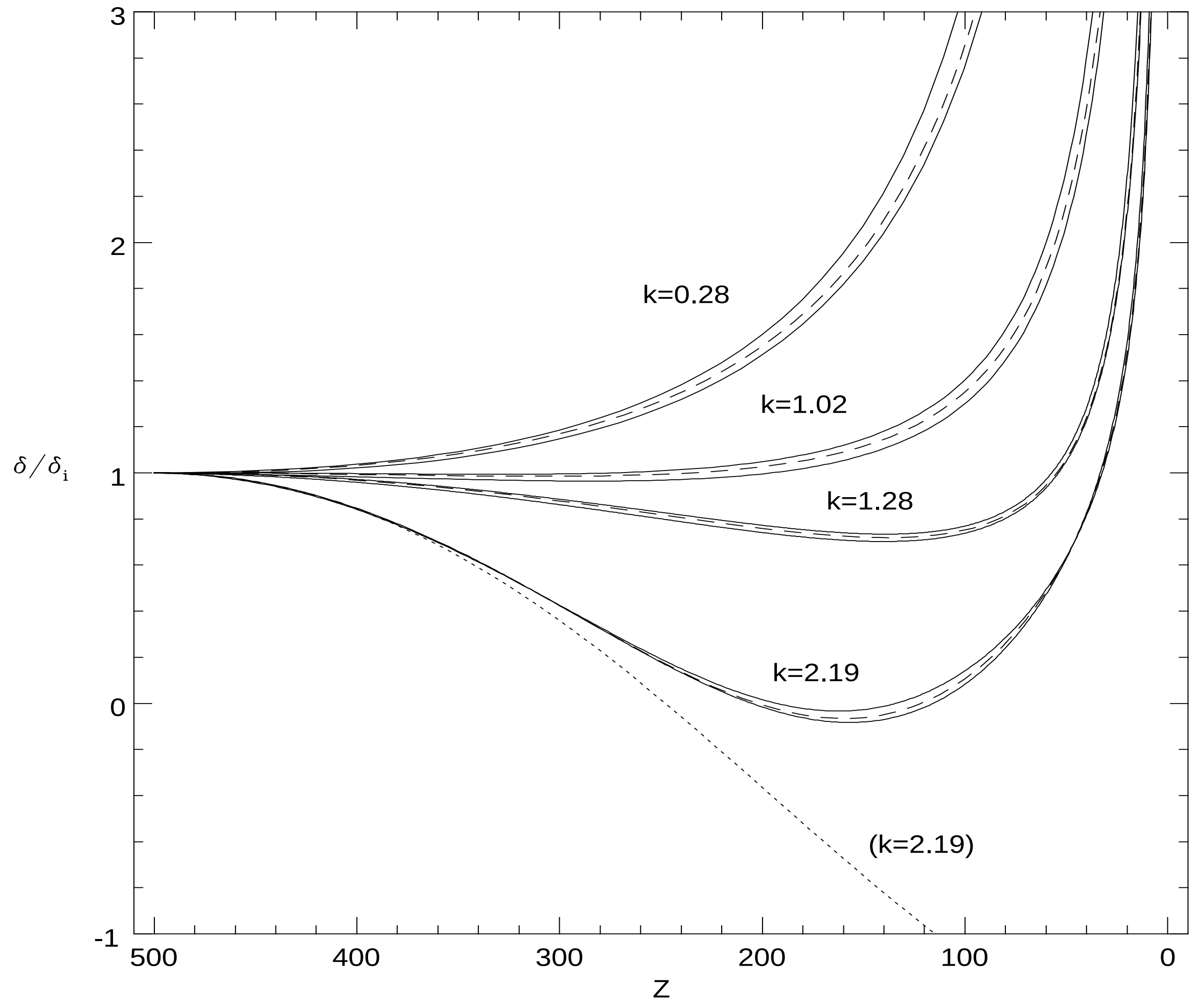

Fig. 1.— Evolution of the overdensity for single spherical Fourier modes in the linear regime. The overdensity $\delta$ is normalized to its initial value $\delta_i$. The comoving wavenumber $k$ is given in $\mathrm{kpc}^{-1}$. The dashed lines are obtained analytically from linear theory. The pairs of solid lines show numerical results for $\delta(r, z)$ at the center and at the surface of the sphere. The overdensity at the center of the sphere becomes slightly higher than at the surface. The dotted line shows the analytic results from linear theory when the temperature gradient term in the square brackets in equation (5) of § 2 is ignored.

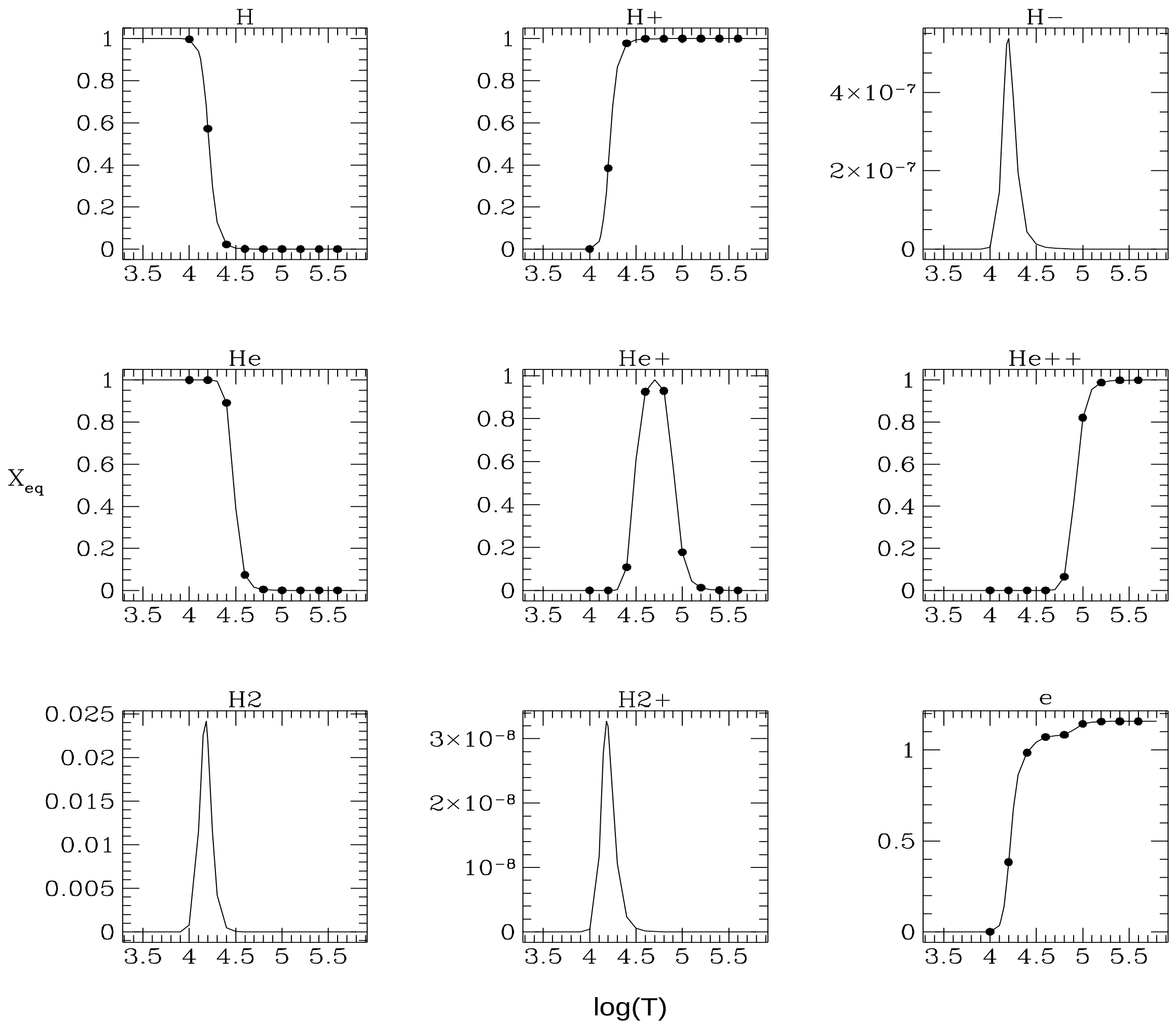

Fig. 2.— Equilibrium abundances of the nine species as a function of temperature. The abundances were obtained by integrating equations (12) with the routine *stifbs* until the abundances reached their equilibrium values. The dots show values obtained algebraically from equations (12) by setting $n_{\mathrm{H}_2}$=$n_{\mathrm{H}^-}$=$n_{\mathrm{H}_2^+}$=0 and $d/dt$=0.

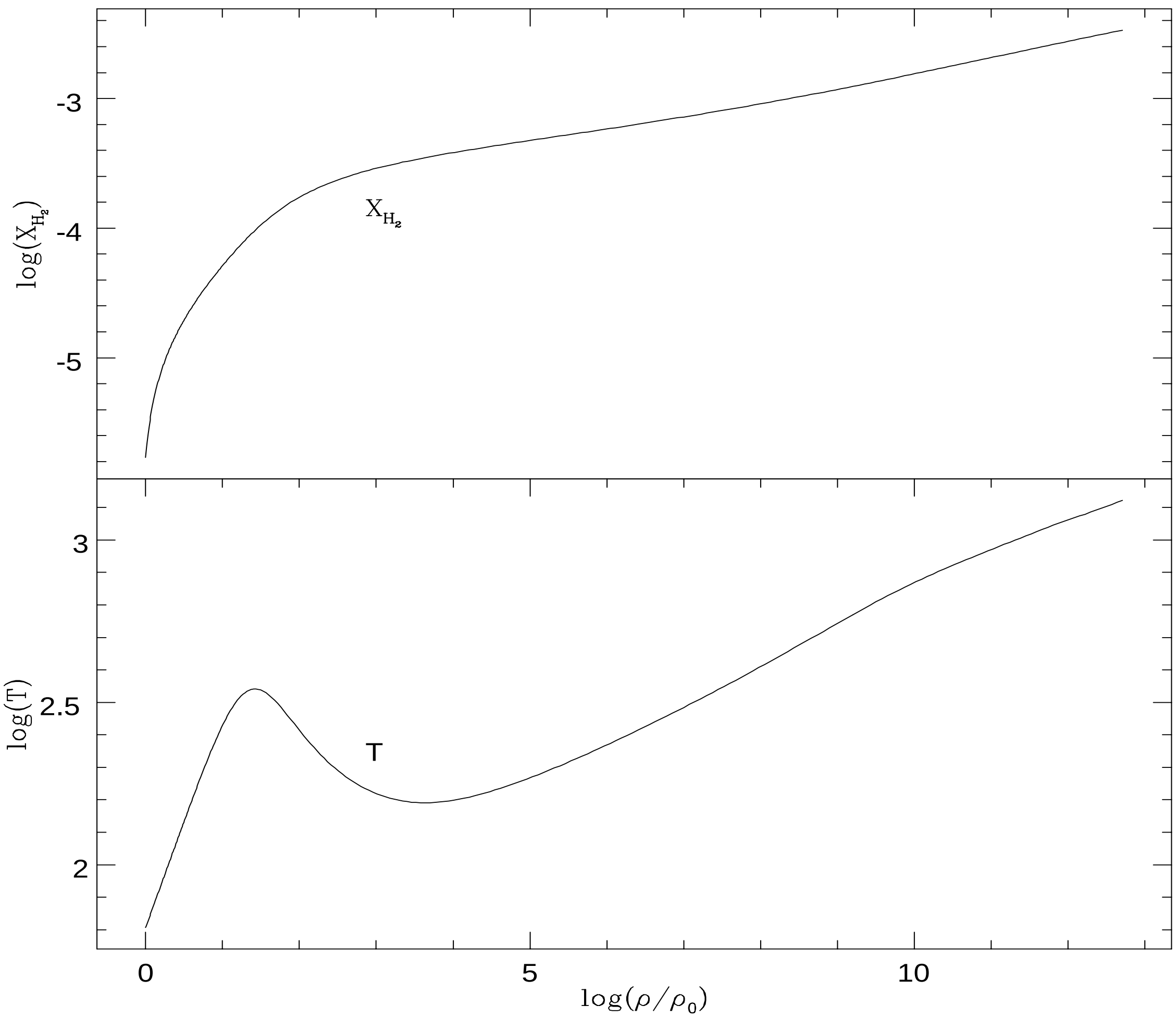

Fig. 3.— Temperature (in K) and fractional abundance of $H_2$ molecules vs. density in a freely collapsing uniform sphere in the presence of $H_2$ cooling. Parameters of the sphere are described in § 3.2(c).

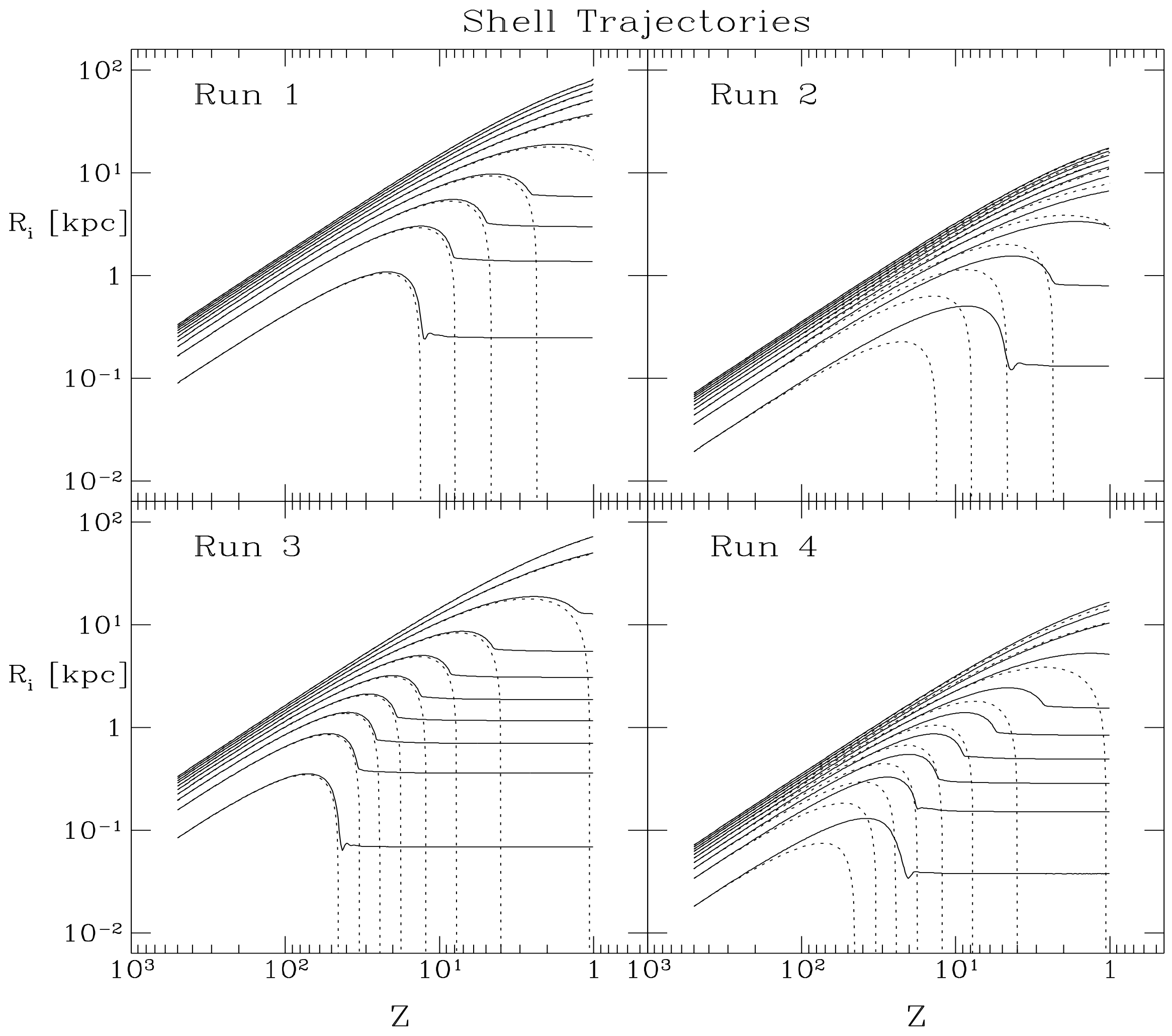

Fig. 4.— Evolution of the radii of 10 gas shells (enclosing 7, 17, 27, ..., 87, and 97% of the total baryonic mass $M_{\rm bound}$) for four pure baryonic runs without cooling. The parameters of these runs are summarized in Table 2. The solid lines show the shell trajectories from the simulations, while the dashed lines show for reference the corresponding shell trajectories for a pressureless gas.

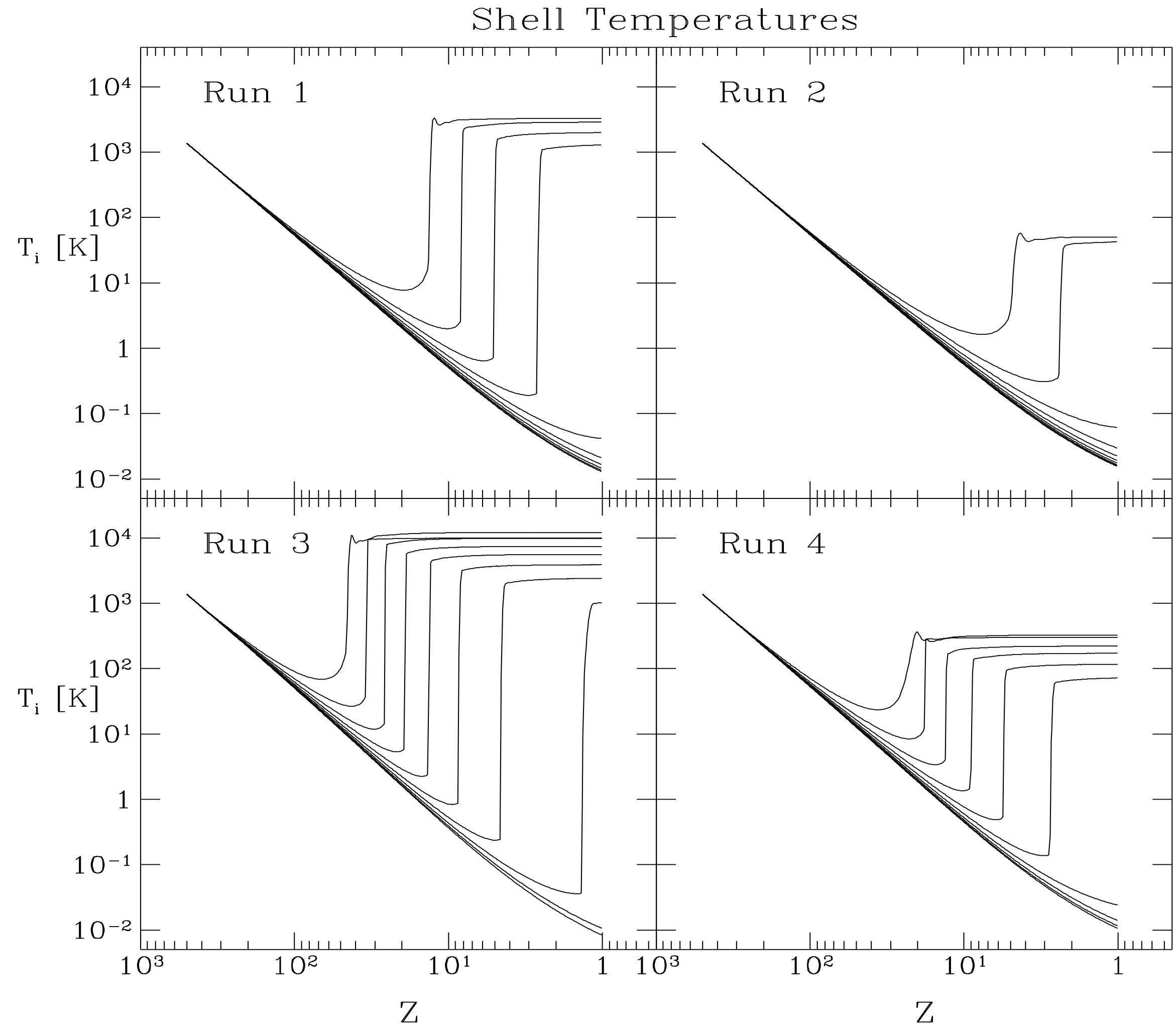

Fig. 5.— Temperature evolution of the gas shells that appear in Figure 4.

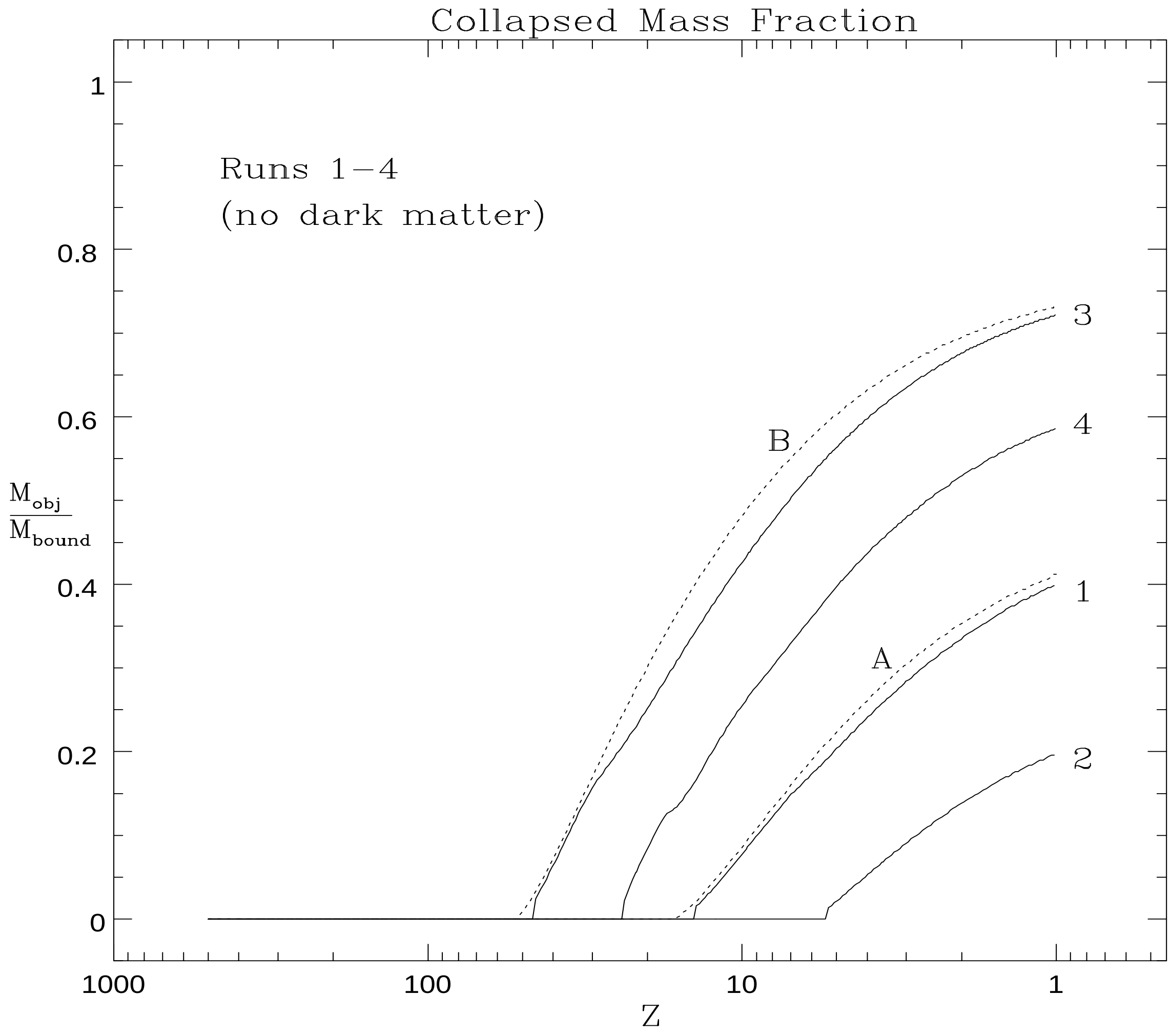

Fig. 6.— The solid lines show the fraction of the bound mass that has collapsed as a function of redshift for the four runs 1-4 in Table 2. The two dashed lines show what the collapsed baryonic mass fraction would be in each case if the cloud had been in free-fall. The line marked "A" is for $\delta_{\rm dm,i} = 0.13$ (to be compared with runs 1 and 2) and the line marked "B" is for $\delta_{\rm dm,i} = 0.39$ (to be compared with runs 3 and 4).

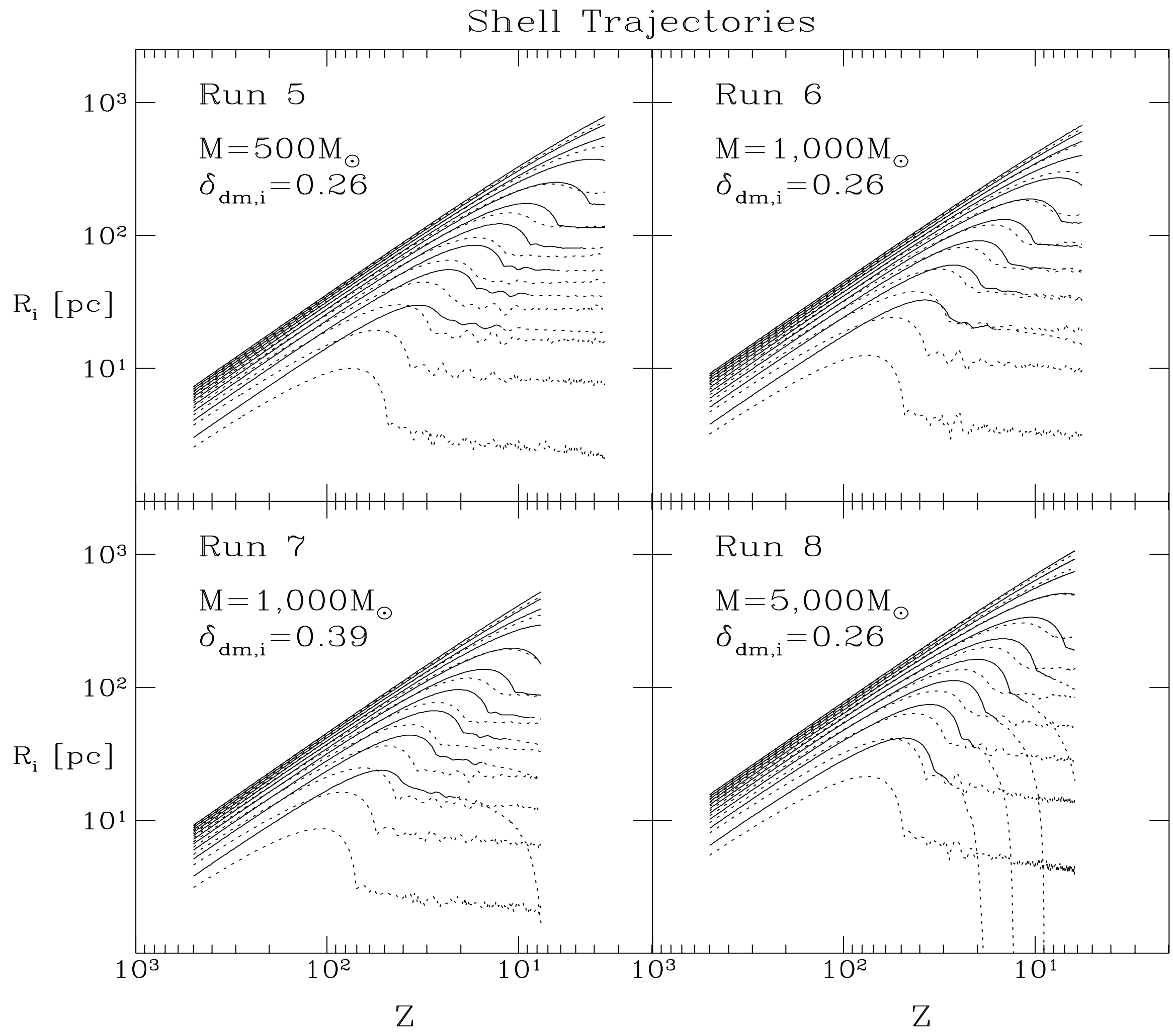

Fig. 7.— Evolution of the radii of 10 gas shells (enclosing 7, 17, 27, ..., 87, and 97% of the total baryonic mass $M_{\rm bound}$) for four mixed baryons and dark matter runs with cooling. The parameters of these runs, numbered 5-8, are summarized in Table 2. The trajectories of dark matter shells are shown as dashed lines. The trajectories of the gas shells are shown as solid lines until they reach $\bar{\delta} = \bar{\delta}_{\rm thr}$, and as dotted lines afterwards.

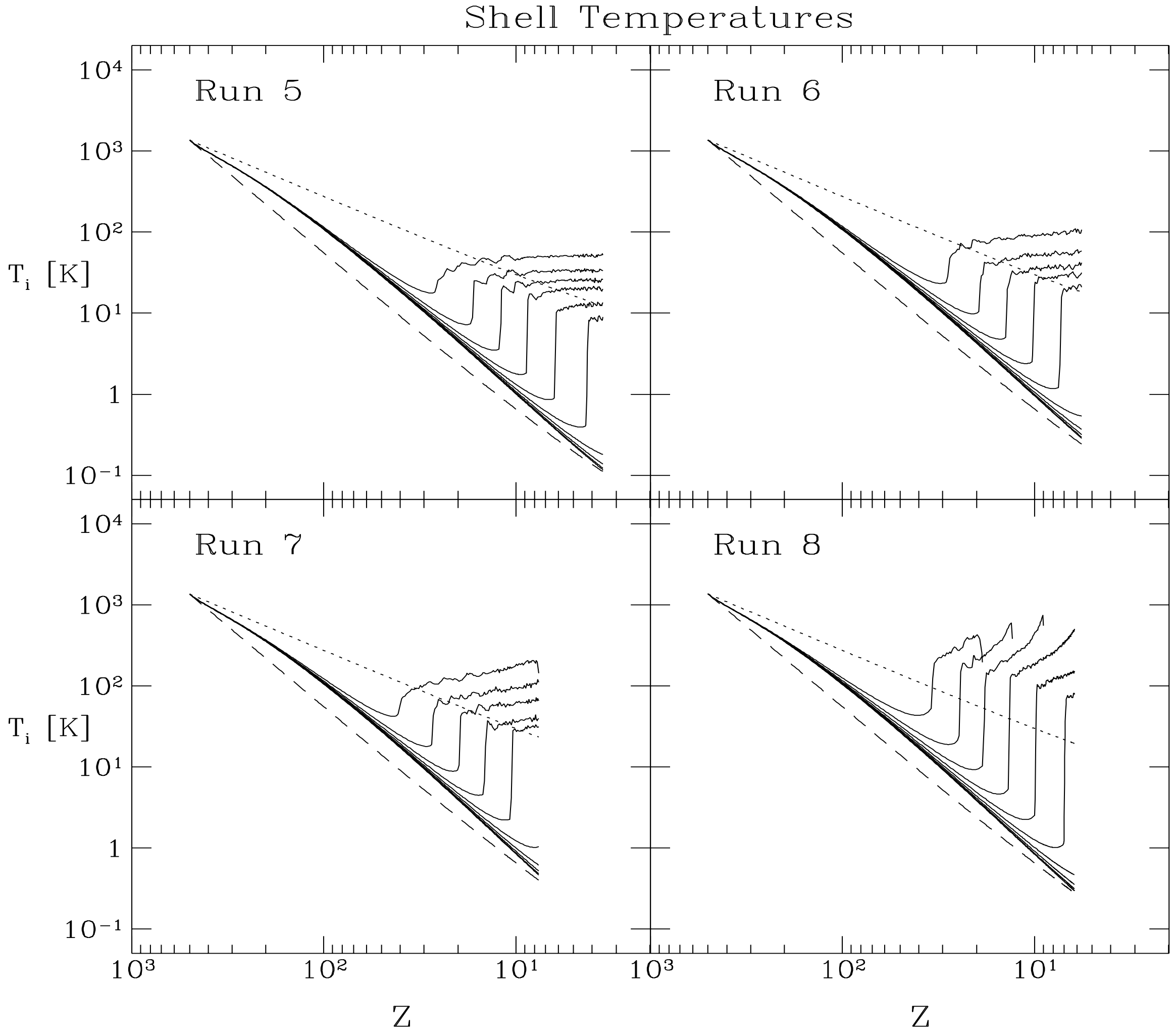

Fig. 8.— Temperature evolution of the gas shells that appear in Figure 7. The dotted lines show the CBR temperature $T_\gamma \propto (1+z)$, and the dashed lines show the adiabatic temperature $T_{\rm ad} \propto (1+z)^2$. The gas temperature remains between $T_\gamma$ and $T_{\rm ad}$ until it shocks.

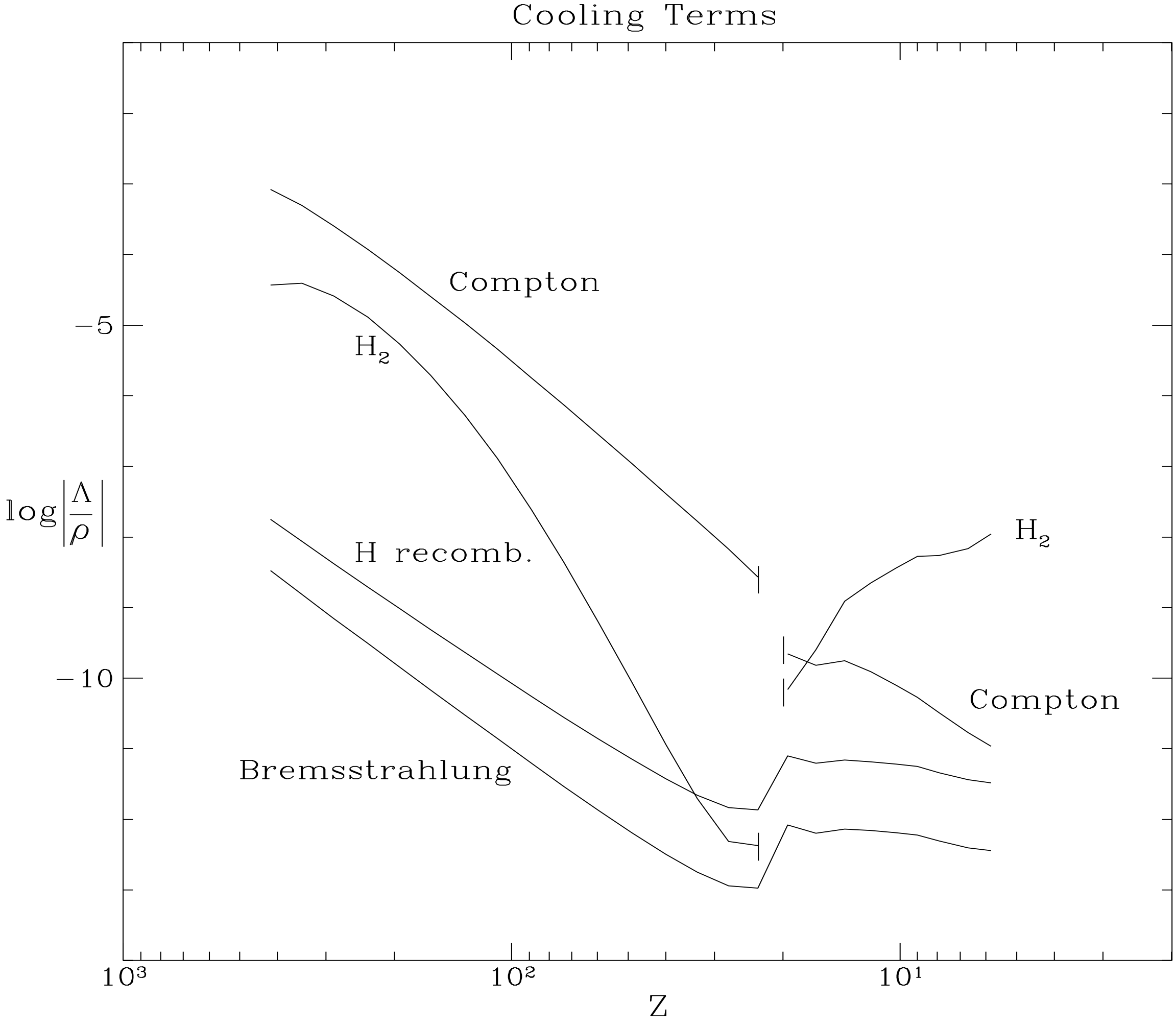

Fig. 9.— The redshift evolution of the four most efficient radiative cooling/heating terms within the shell enclosing 12% of the total baryonic mass $M_{\rm bound}$ in a typical run (number 6). Before the gas temperature rises above the CBR temperature, both Compton scattering and the radiative transitions of $H_2$ heat the gas. $\Lambda$ changes sign for these processes in-between the tick marks. After $T_{\rm b} > T_\gamma$ (or $z \lesssim 22$), both of these mechanisms cool the gas; the most efficient post-shock cooling is due to $H_2$. The units of $\Lambda/\rho$ are erg $s^{-1}g^{-1}$.

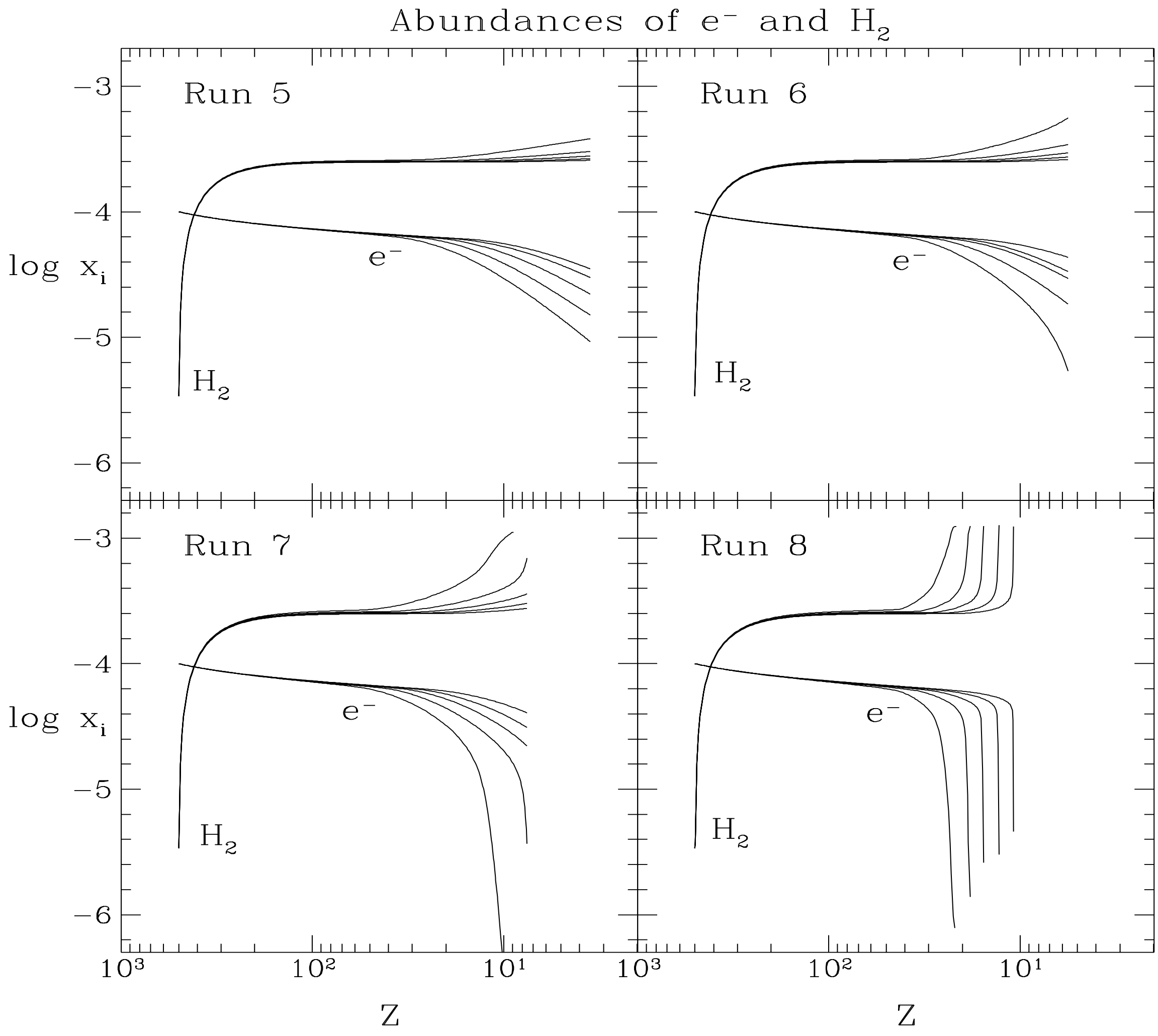

Fig. 10.— Evolution of the relative abundances of $H_2$ and $e^-$ for 5 shells (enclosing 2, 7, 12, 17, and 23% of the total baryonic mass $M_{\rm bound}$) in the runs numbered 5-8.

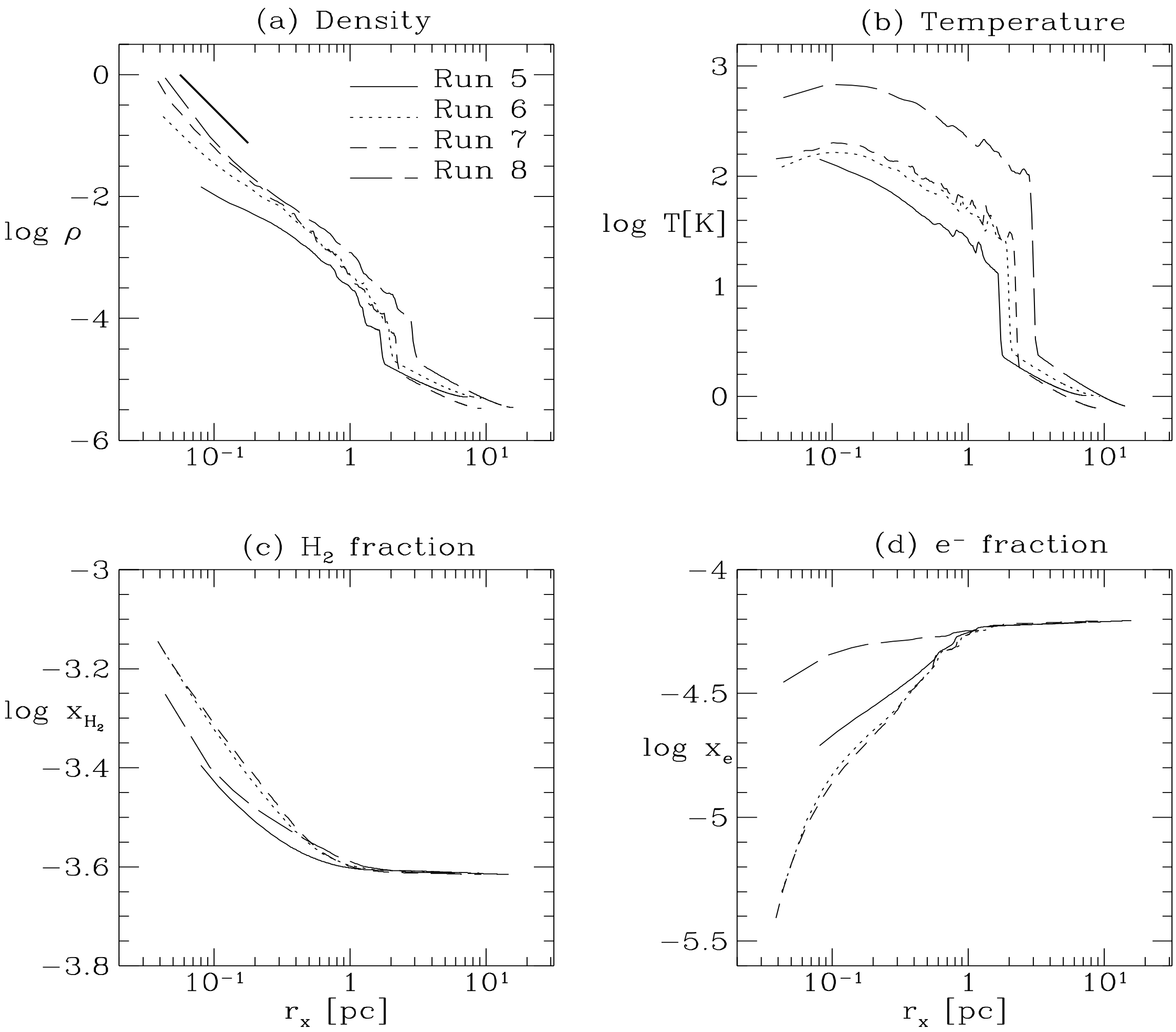

Fig. 11.— Density, temperature, relative $H_2$ abundance and $e^-$ abundance profiles in runs 5-8 at redshift $z = 10$. The abscissa is the comoving radius $r_x = r(1 + 500)/(1 + z)$, where $r$ is the physical radius. The density $\rho$ is given in units of $M_\odot/\mathrm{pc}^3$.
The thick line on the top left corner of panel (a) shows the slope of the density profile in the pressureless self-similar solution (Bertschinger 1985).

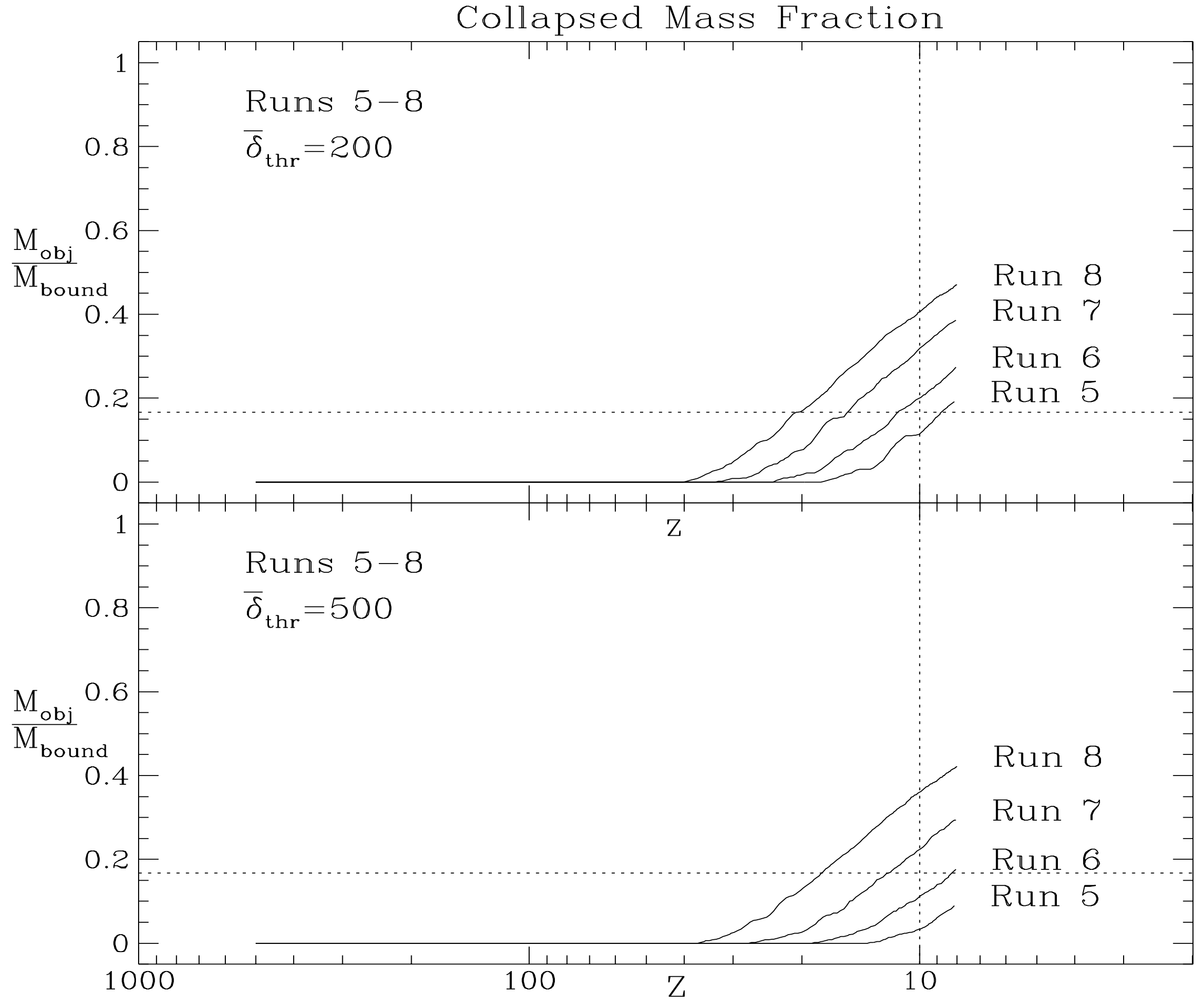

Fig. 12.— Fraction of the total baryonic bound mass that has collapsed and formed an object in runs 5-8 vs. redshift. The upper panel shows results with $\bar{\delta}_{\rm thr} = 200$, the lower panel with $\bar{\delta}_{\rm thr} = 500$. The vertical and horizontal dashed lines illustrate the condition $M_{\rm obj}/M_{\rm cloud} \geq 0.5$ (or equivalently $M_{\rm obj}/M_{\rm bound} \geq 0.17$) at $z = 10$.

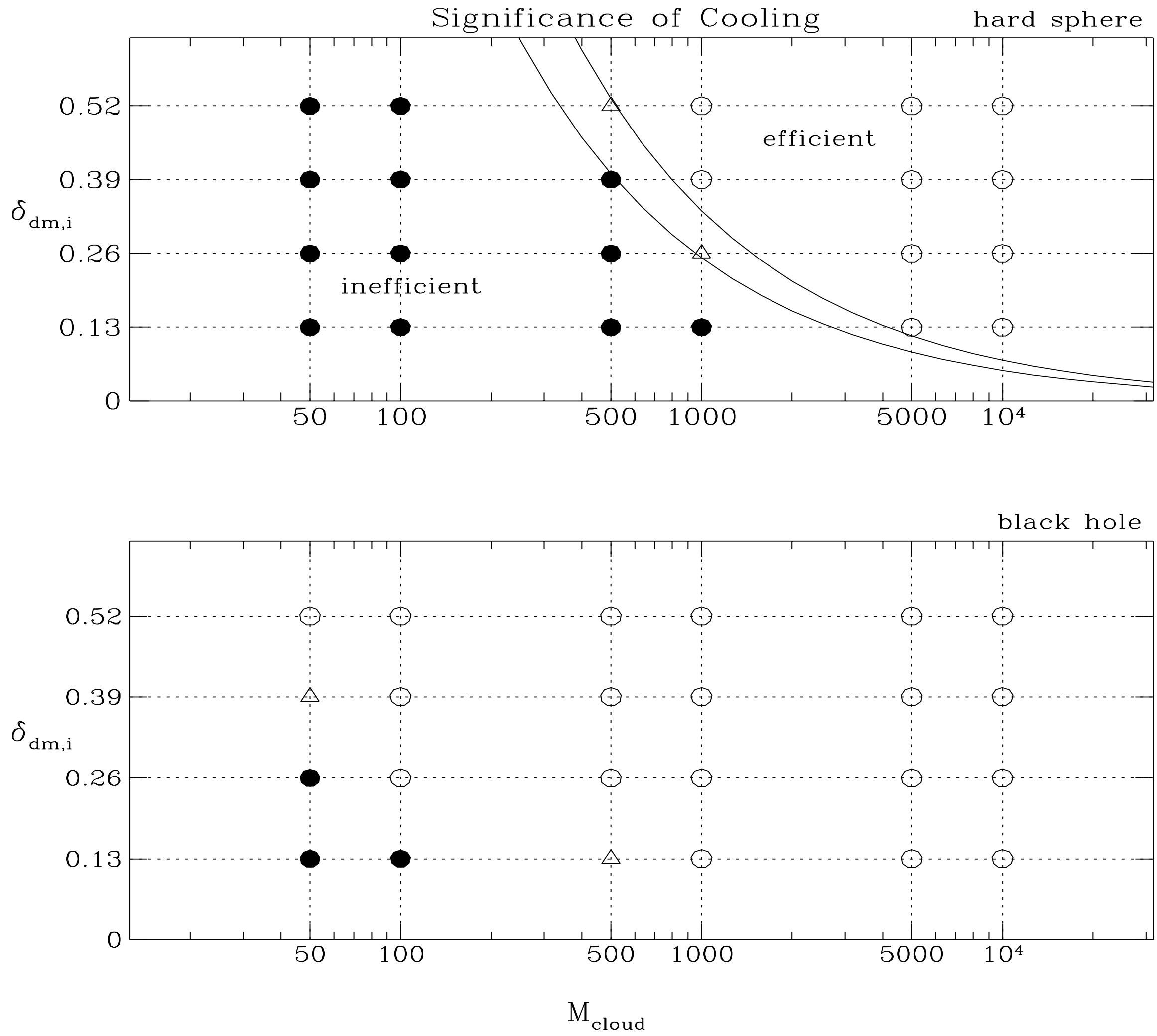

Fig. 13.— Effectiveness of radiative cooling as a function of the two parameters $M_{cloud}$ and $\delta_{dm,i}$. Filled circles represent cases where cooling had no effect; open circles denote cases where cooling was very efficient; open triangles represent intermediate cases where cooling had minor but noticeable effect. On these mass-scales, $\delta_{dm,i} = 0.13$ corresponds to (1.1-1.4)$\sigma$ for the COBE-normalized ($\sigma_8 = 1.22$) cold dark matter power-spectrum (cf. Table 1). The upper panel refers to runs with a reflecting sphere in the center, and the lower panel to runs with a central black hole. The solid curves are lines of constant virial temperature (cf. eq. 19); the upper and lower curves correspond to temperatures of 120 K and 85 K, respectively.

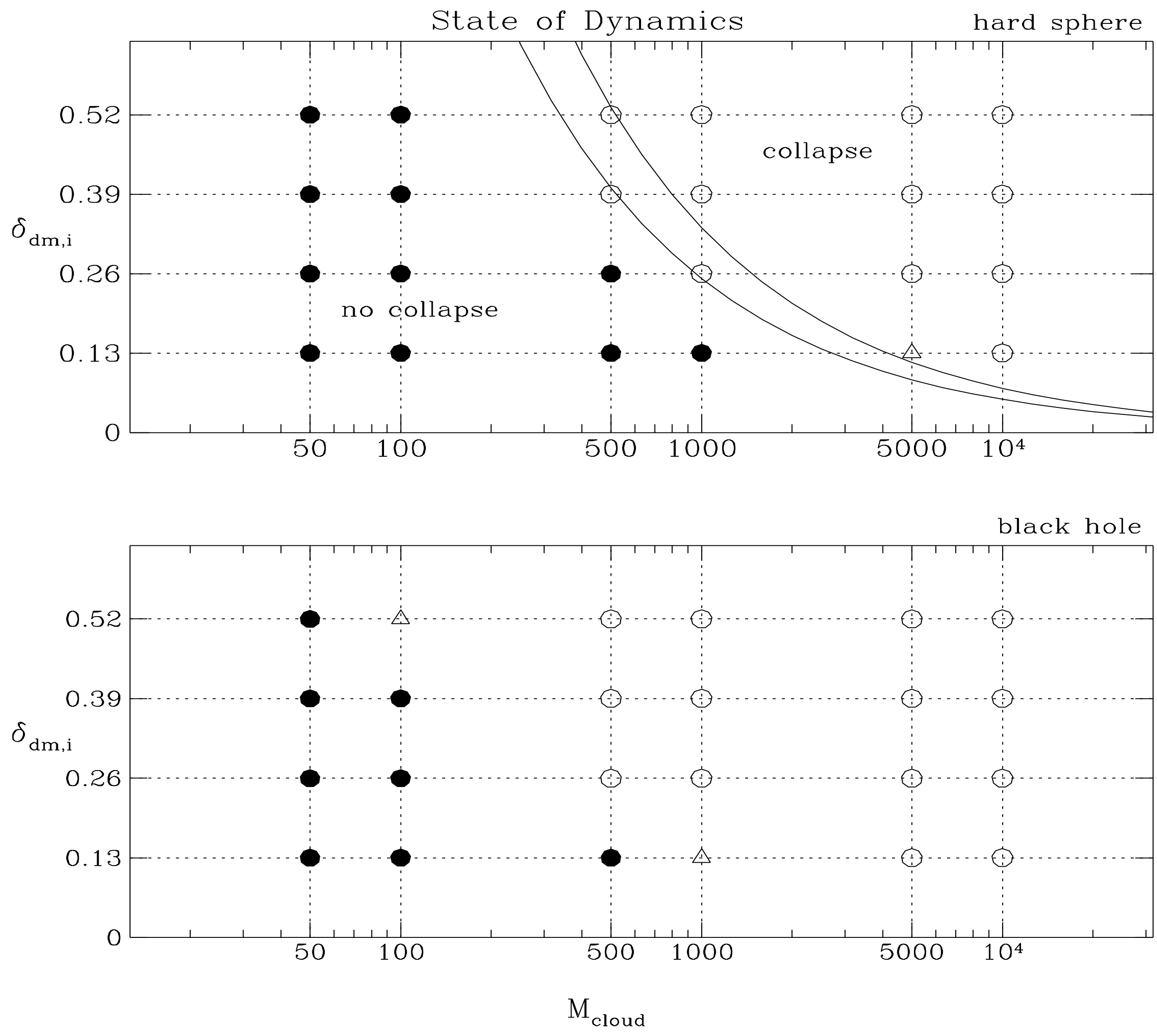

Fig. 14.— Collapse as a function of the two parameters $M_{cloud}$ and $\delta_{dm,i}$. The definition of "collapse" is given in § 4.2. Filled circles represent runs where none of the shells have collapsed by the redshift $z = 10$; open circles represent runs where more than half of the baryonic mass $M_{cloud}$ has collapsed by $z = 10$; open triangles represent intermediate cases where only the innermost few shells have collapsed. The upper panel refers to runs with a reflecting sphere in the center, and the lower panel to runs with a central black hole. The solid lines are curves of constant virial temperature, as in Figure 13.